\pdfoutput=1
\documentclass[acmlarge]{acmart}

\usepackage{booktabs} 
\usepackage{color}
\usepackage{multirow}
\usepackage{graphicx}
\usepackage{xspace}
\usepackage{subfigure}  
\usepackage{epsfig}
\usepackage{epstopdf}
\usepackage{bm}
\usepackage{amsmath}
\usepackage{amssymb}
\usepackage{url}
\usepackage{array}
\usepackage{amsfonts}

\usepackage[ruled]{algorithm2e} 

\SetAlFnt{\small}
\SetAlCapFnt{\small}
\SetAlCapNameFnt{\small}
\SetAlCapHSkip{0pt}
\IncMargin{-\parindent}

\makeatletter

\newcommand{\Rmnum}[1]{\expandafter\@slowromancap\romannumeral #1@}
\makeatother

\acmJournal{TOPS}

\newif\ifdebugdoc\debugdoctrue

\ifdebugdoc

\newcommand{\scott}[1]{\textcolor{red}{\textbf{(scott: #1)}}}
\newcommand{\outline}[1]{\textbf{\colorbox{yellow}{Outline:}\textcolor{red}{#1.}}}
\newcommand{\hl}[1]{\textcolor{black}{{#1}}}
\newcommand{\fyi}[1]{\footnote{\textcolor{blue}{fyi:#1}}}
\newcommand{\todo}[1]{{\colorbox{yellow}{[TODO:}\textcolor{blue}{#1}]}}

\newcommand{\add}[1]{\textcolor{red}{#1}}
\newcommand{\del}[1]{\textcolor{blue}{\sout{#1}}}

\newcommand{\fdel}[1]{\textcolor{blue}{\sout{#1}}}

\else
\newcommand{\scott}[1]{}
\newcommand{\outline}[1]{}

\newcommand{\hl}[1]{\textcolor{blue}{#1}}
\newcommand{\fyi}[1]{}
\newcommand{\todo}[1]{}

\newcommand{\add}[1]{#1}
\newcommand{\del}[1]{}

\newcommand{\fdel}[1]{}
\fi

\begin{document}
	\title{The Untold Secrets of Operational Wi-Fi Calling Services: Vulnerabilities, Attacks, and Countermeasures}
	
	\author{Tian Xie}
	\orcid{1234-5678-9012-3456}
	\affiliation{%
		\institution{Michigan State University}
		\streetaddress{428 S Shaw Ln}
		\city{East Lansing}
		\state{MI}
		\postcode{48823}}
	\email{xietian1@msu.edu}

	\author{Guan-Hua Tu}
	\affiliation{%
		\institution{Michigan State University}
		\city{East Lansing}
	}
	\email{ghtu@msu.edu}
		
	\author{Bangjie Yin}
	\orcid{1234-5678-9012-3456}
	\affiliation{%
	\institution{Michigan State University}
	\streetaddress{428 S Shaw Ln}
	\city{East Lansing}
	\state{MI}
	\postcode{48823}}
	\email{yinbangjie1@msu.edu}

	\author{Chi-Yu Li}
	\affiliation{%
		\institution{National Chiao Tung University}}
	\email{chiyuli@cs.nctu.edu.tw}
	\author{Chunyi Peng}
	\affiliation{%
		\institution{Purdue University}
	}
	\email{chunyi@purdue.edu}
	\author{Mi Zhang}
	\affiliation{%
		\institution{Michigan State University}
		\city{East Lansing}
	}
	\email{mizhang@msu.edu}
	\author{Hui Liu}
	\orcid{1234-5678-9012-3456}
	\affiliation{%
		\institution{Michigan State University}
		\streetaddress{428 S Shaw Ln}
		\city{East Lansing}
		\state{MI}
		\postcode{48823}}
	\email{liuhui@msu.edu}
		\author{Xiaoming Liu}
	\affiliation{%
		\institution{Michigan State University}
		\city{East Lansing}
	}
	\email{liuxiaoming@msu.edu}

	\begin{abstract}
		
		Since 2016, all of four major U.S. operators have rolled out nationwide Wi-Fi calling services.
		They are projected to surpass VoLTE (Voice over LTE) and other VoIP services in terms of mobile IP voice usage minutes in 2018.
		They enable mobile users to place cellular calls over Wi-Fi networks based on the 3GPP IMS (IP Multimedia Subsystem) technology.
		Compared with conventional cellular voice solutions, the major difference lies in that their traffic traverses untrustful Wi-Fi networks and the Internet.
		This exposure to insecure networks may cause the Wi-Fi calling users to suffer from security threats.
		Its security mechanisms are similar to the VoLTE, because both of them are supported by the IMS.
		They include SIM-based security, 3GPP AKA (Authentication and Key Agreement), IPSec (Internet Protocol Security), etc.
		However, are they sufficient to secure Wi-Fi calling services?
		Unfortunately, our study yields a negative answer. 
		We conduct the first study of exploring security issues of the operational Wi-Fi calling services in three major U.S. operators' networks using commodity devices.
		We disclose that current Wi-Fi calling security is not bullet-proof and uncover four vulnerabilities which stem from improper standard designs, device implementation issues and network operation slips.
		By exploiting the vulnerabilities, together with several state-of-the-art computer visual recognition technologies, we devise two proof-of-concept attacks: user privacy leakage and telephony harassment or denial of voice service (THDoS); both of them can bypass the security defenses deployed on mobile devices and the network infrastructure.
		We have confirmed their feasibility and simplicity using real-world experiments, as well as assessed their potential damages and proposed recommended solutions.
	\end{abstract}

	%
	%

	
	%
	%

	
	 \authorsaddresses{Extension of Conference Paper: a preliminary version of this article entitled `The Dark Side of Operational Wi-Fi Calling Services' by T. Xie et al. appeared in IEEE Conference on Communications and Network Security (IEEE CNS), 2018.\\
 	Authors' addresses: T.~Xie, G.~H.~Tu, B.~J.~Yin, Mi~Zhang, Hui~Liu, X.~M.~Liu, Department of Computer Science and Engineering, Michigan State University, East Lansing, MI, USA, 48823. Chi-Yu Li, Department of Computer Science, National Chiao Tung University, Taiwan. Chunyi~Peng, Department of Computer Science, Purdue University, West Lafayette, IN, 47907.}
	
	\maketitle
	
	\renewcommand{\shortauthors}{T. Xie et al.}

	\section{Introduction}




Since 2016, all of four major operators in the U.S., T-Mobile, AT\&T, Verizon and Sprint, have launched their nationwide Wi-Fi calling services~\cite{GSMA-WiFi-Calling}\footnote{It is also named as VoWiFi (Voice over Wi-Fi).}. The Wi-Fi calling technology utilizes the 3GPP IMS (IP Multimedia Subsystem) system~\cite{TS23.237} to provide a packet-switched voice service over Wi-Fi networks. It enables mobile users to have cellular calls and text messages
through their home/public Wi-Fi networks instead of cellular base stations. It is considered as an 
alternative voice solution for the mobile users with weak signals of base stations.
A recent Cisco report~\cite{WiFi-calling-forecast-2016} forecasts that the Wi-Fi calling is going to surpass VoLTE (Voice over LTE) and VoIP (Voice over IP, e.g., Microsoft Skype and Google Hangouts) services  by 2018 in terms of voice usage minutes. By 2020, the Wi-Fi calling will take 53\% of mobile IP voice service usage (about 9,000 billions of minutes per year), whereas the VoLTE and other VoIP services will have only 26\% and 21\%, respectively.
As a result, any security loopholes of the Wi-Fi calling can lead to devastating consequences on a global scale due to its rapid global deployment~\cite{wifi_global_apple}. We believe that there is a critical need for a security investigation on Wi-Fi calling services.

Technically, Wi-Fi calling services differ from the proprietary VoIP services such as Skype or other SIP-based (Session Initiation Protocol) voice services.
Though its signaling protocol is also SIP-based, it is a 3GPP-specific version~\cite{rfc4083,rfc7315}. For security reasons, both 3GPP and GSMA stipulate that Wi-Fi calling services shall use well-examined SIM-based security (i.e., storing each user's private secret key in a physical card) and authentication methods (i.e., 3GPP AKA (Authentication and Key Agreement)~\cite{TS33.401}), which are employed by the VoLTE.
In addition, all the Wi-Fi calling signaling and voice/text packets shall be delivered through IPSec (Internet Protocol Security) channels between Wi-Fi calling devices and the cellular network infrastructure, since they may cross public, insecure networks. To defend against Wi-Fi DoS (Denial-of-Service) attacks (e.g., all the Wi-Fi calling packets are discarded by malicious Wi-Fi networks), the Wi-Fi calling employs a system-switch security mechanism, which switches Wi-Fi calling users back to cellular-based voice/text services when the users are unreachable through Wi-Fi networks.


Given these security mechanisms, which have been well studied in the VoLTE~\cite{li2015insecurity} and cellular networks~\cite{bhattarai2014simulation} for years, it seems that the Wi-Fi calling should be as secure as the VoLTE. Unfortunately, it is not the case. We have identified several security threats in the Wi-Fi calling services deployed by T-Mobile, Verizon and AT\&T in the U.S.
The threats can be attributed to design defects of Wi-Fi calling standards, implementation issues of Wi-Fi calling devices, and operational slips of cellular networks.
Specifically, we discover four vulnerabilities. First, Wi-Fi calling devices do not exclude insecure Wi-Fi networks which may impede their Wi-Fi calling services from their selection (V1). Second, they do not defend against ARP spoofing/poisoning attacks, which can be exploited to launch various MITM (Man-In-The-Middle) attacks (V2).
Third, the Wi-Fi calling traffic, which is protected by the IPSec, is still vulnerable to side-channel attacks (e.g., privacy leakage) (V3). 
Fourth, even when the performance of a Wi-Fi calling call is bad (e.g., voice is muted), the mechanism of service continuity across the Wi-Fi calling and the cellular-based voice services does not take effect (V4). 
They are summarized in Table~\ref{tab:summary-vul}.

\begin{table*}
	\centering
	\scriptsize
	\resizebox{1\textwidth}{!}{
		\small
		\begin{tabular}{|p{2cm}|p{6cm}|p{2cm}|p{9cm}|}
			\hline
			\textbf{Category} & \textbf{Vulnerability} & \textbf{Type} & \textbf{Root Cause}\\
			\hline
			\hline
			\multirow{8}{*}{\textbf{Device}}  & \multirow{4}{*}{\parbox{6cm}{V1: Wi-Fi calling devices do not exclude insecure Wi-Fi networks from their selection.}}  & \multirow{4}{*}{Design defect}  & \multirow{4}{*}{\parbox{9cm}{Current 3GPP Wi-Fi network selection mechanism~\cite{GSMA-WiFi-Calling,TS24.302} considers only the connectivity capabilities of Wi-Fi networks, but not their security risks (Section~\ref{subsubsect:q1}).}} \\
			&&&\\
			&&&\\
			&&&\\
			\cline{2-4}
			& \multirow{3}{*}{\parbox{6cm}{V2: Wi-Fi calling devices do not defend against ARP spoofing/poisoning attacks.}}  & \multirow{3}{*}{\parbox{2cm}{Implementation issue}}  & \multirow{3}{*}{\parbox{9cm}{Wi-Fi calling devices do not advance the Wi-Fi security for Wi-Fi calling services (Section~\ref{subsubsect:q1}).}} \\
			&&&\\
			&&&\\
			\hline
			\multirow{9}{*}{\textbf{Infrastructure}}

			& \multirow{5}{*}{\parbox{6cm}{V3: the Wi-Fi calling traffic, which is protected by the IPSec, is still vulnerable to side-channel attacks (e.g., privacy leakage).}}  & \multirow{5}{*}{Operation slip}  & \multirow{5}{*}{\parbox{9cm}{For all three carriers, T-Mobile, AT\&T, and Verizon, the IPSec session between mobile devices and the core network carries only Wi-Fi calling traffic, so traffic patterns can be  learned to infer different events easily.
					(Section~\ref{subsubsect:q1}).}} \\
			&&&\\
			&&&\\
			&&&\\
\cline{2-4}
& \multirow{4}{*}{\parbox{6cm}{V4: service continuity is not updated accordingly.}}
& \multirow{4}{*}{Design defect}  & \multirow{4}{*}{\parbox{9cm}{The triggers of 3GPP SRVCC/DRVCC~\cite{TS23.216,TS23.237,TS24.237} procedures, which keep service continuity across different radio access technologies, consider only radio quality but not service quality (Section~\ref{subsubsect:q2}).}} \\
			&&&\\
			&&&\\
			&&&\\
            &&&\\

			\hline
	\end{tabular}}
    \vspace{0.2cm}
	\caption{Summary of identified security vulnerabilities.}
	\label{tab:summary-vul}
\end{table*}


We exploit these four vulnerabilities to devise two proof-of-concept attacks: (1) user privacy leakage and (2) telephony harassment or denial of voice service attack (THDoS).
They can bypass the existing security mechanisms on the Wi-Fi calling devices and the network infrastructure. Note that in our threat model, the adversary has no control over the Wi-Fi network used by the victim, the victim's Wi-Fi calling device, or the cellular network infrastructure.

For the user privacy leakage, we develop a Wi-Fi calling user privacy inference system (UIPS), which employs several state-of-the-art computer visual recognition technologies. It can infer a verify of user privacy including Wi-Fi user identifiers (e.g., names), call statistics, Wi-Fi calling device addresses (i.e., IP addresses), device information (e.g., Apple iPhone8), user activities (e.g., accessing CNN.com), and applications running on the device (e.g., Wechat). Note that call statistics have been proven effective to infer a user's personality (e.g., conscientiousness~\cite{de2013predicting}), mood (e.g., stressful~\cite{thomee2011mobile}), and behavior (e.g., dialing spamming calls~\cite{DBLP:conf/ceas/BalasubramaniyanAP07}). For the THDoS attack, we devise four attack variances against Wi-Fi calling users: annoying-incoming-call, zombie-call, mute voice call, and telephony denial-of-voice-service. We further propose solutions to address these security threats.

In summary, this paper makes three main contributions.
\begin{enumerate}
  \item We conduct the first security study to explore the dark side of operational Wi-Fi calling services in three major U.S. operators' networks (i.e., T-Mobile, AT\&T, and Verizon) using commodity devices. We identify four vulnerabilities which root in design defects of Wi-Fi calling standards, operational slips of operators, and implementation issues of Wi-Fi calling devices.
  \item We devise two proof-of-concept attacks by exploiting the identified vulnerabilities, together with several computer visual recognition technologies. We further assess their impacts in those three U.S. carriers' networks.
  \item We identify diversified root causes and propose recommended solutions. The lessons learned can facilitate and secure the global deployment of Wi-Fi calling services.
\end{enumerate}



The rest of the paper is organized as follows. Section~\ref{sect:what_may_go_wrong} presents \hl{the background of Wi-Fi calling technology and possible security threats, as well as our threat model and methodology.} 
Section~\ref{sect:how_does_it_go_wrong} discloses the discovered vulnerabilities of Wi-Fi calling services.
Sections~\ref{sect:privacy_leakage} and ~\ref{sect:thdos} show the proof-of-concept attacks, \hl{user privacy leakage and the THDoS}, respectively. 
We propose recommended solutions in Section~\ref{sect:sol} and review related work in Section~\ref{sect:related}. Section~\ref{sect:concl} concludes this paper.

	\section{WHAT MAY PROBABLY GO WRONG?}
\label{sect:what_may_go_wrong}

In this section, \hl{we introduce the support of the Wi-Fi calling service in cellular networks, and its possible security threats.
We then present the threat model and the experimental methodology of this work.}

\subsection{Wi-Fi calling Primer} The Wi-Fi calling technology 
enables mobile users to consume cellular-based voice services through Wi-Fi networks.
\hl{It is an alternative voice solution for the conditions of poor cellular signals, which can be caused by insufficient coverage from limited base stations.}
\hl{By design,} it is similar to most VoIP (Voice over IP) services by using the SIP (Session Initiation Protocol) as its signaling protocol.
\hl{But, its SIP requires some 3GPP-specific modifications (\cite{rfc4083,rfc7315}) and it is supported by the cellular core network.}



\begin{figure}[t]
	\includegraphics[width=0.8\columnwidth]{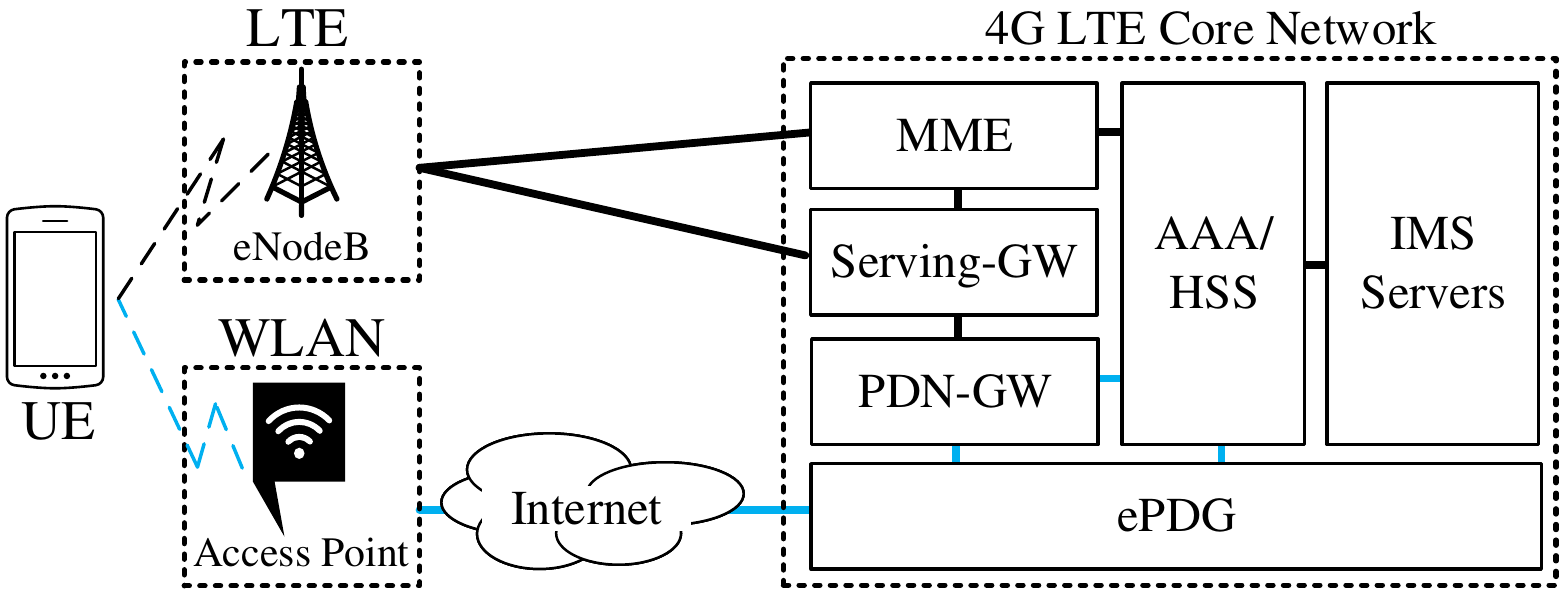}
	\caption{\hl{The 4G LTE network architecture that supports the Wi-Fi calling service.}}
	\label{fig:volte_vowifi}
\end{figure}

Figure~\ref{fig:volte_vowifi} illustrates a simplified network architecture \hl{that runs both Wi-Fi calling and VoLTE services}. 
Both Wi-Fi calling and VoLTE users use the same infrastructure consisting of two parts: Radio Access Network (RAN) and Core Network (CN). Unlike \hl{the VoLTE that} leverages eNodeB as its RAN, the Wi-Fi calling adopts the Wi-Fi Access Point (AP), \hl{which forwards its packets to the CN through the Internet.}
\hl{The UE (User Equipment) can dynamically choose the Wi-Fi calling or the VoLTE as its voice service.}
The CN consists of six main components: the ePDG (Evolved Packet Data Gateway), the PDN-GW (Public Data Network Gateway), the MME (Mobility Management Entity), the S-GW (Servicing Gateway), the AAA (Authentication, Authorization, and Accounting) server, and the IMS (IP Multimedia Subsystem) servers. The ePDG is newly deployed for the Wi-Fi calling service, \hl{whereas the other components are shared by the Wi-Fi calling and the VoLTE.}

\hl{For the security of the Wi-Fi calling service, each UE has to be authenticated by the ePDG with the AAA server's assistance.
Afterwards, the ePDG establishes an IPSec (Internet Protocol Security) tunnel with the UE for their secure communication.
The ePDG then forwards packets} between the UE, the PDN-GW and the IMS servers, which offer the Wi-Fi calling service.


\subsection{Possible Security Threats}
\hl{The radio access of the Wi-Fi calling technology relies on the Wi-Fi, but not the cellular base stations fully controlled by operators.} It naturally raises \hl{some} security concerns. 
\emph{Is the Wi-Fi calling technology as secure as the traditional voice service?} In particular, we start with the following two questions.

\begin{itemize}
	\item[Q1.] Do Wi-Fi calling devices still activate Wi-Fi calling services while associating with insecure Wi-Fi networks?
	\item[Q2.] If yes, do any security mechanisms exist on the devices or/and the network infrastructure to defend against \hl{possible} security threats? Moreover, can \hl{they} be completely eliminated?
\end{itemize}



Unfortunately, we disclose that operational Wi-Fi calling services, as well as their technical supports behind, are not bullet-proof.
Wi-Fi calling devices do not avoid \hl{activating} Wi-Fi calling services in insecure Wi-Fi networks (Q1); although the devices and the infrastructure provide security mechanisms against some malicious attacks (e.g., DoS attacks),
they do not defend against all the threats(Q2). 
We then uncover vulnerabilities from four findings in three aspects: design defects in the standard, operational slips from the operator's network, and the device's implementation issues. 
We elaborate on each finding with its correlative vulnerability in Section~\ref{sect:how_does_it_go_wrong}.


\subsection{Threat model} In this work, the victim is a \hl{mobile user whose device has the Wi-Fi calling service and associates with a Wi-Fi AP.}
The adversary can be anyone who has a networked device under the same subnet as the victim.
S(he) does not need to associate with the Wi-Fi AP that the victim connects.
Take a campus network as an example. The adversary needs to connect to the campus network but can be anywhere on the campus, whereas the victim can \hl{associate with any campus AP}. 
They are under the same subnet of the campus' network gateway.
The adversary does not have any IPSec keys of the victim and is not able to access the victim's device or the cellular network infrastructure. 
The device software/hardware and the infrastructure are not compromised.

\subsection{Methodology}
\hl{We validate vulnerabilities and attacks }
on three major U.S. carriers: T-Mobile, Verizon and AT\&T. They launch Wi-Fi calling services since 2013, 2016 and 2016 respectively.
They together take more than 75\% of market share~\cite{operators-market-share-2017} in the U.S.
We conduct experiments using a software-based Wi-Fi AP on a MacBook Pro 2014 laptop, an ASUS RT-AC1900 Wi-Fi AP, and eight popular mobile devices with Wi-Fi calling services, which include Samsung Galaxy S6/S7/S8/J7, Apple iPhone6/iPhone7/iPhone8, and Google Nexus 6P. Apple and Samsung take 74\% of market share~\cite{smartphone-market}. 
The experiments are done in several campus Wi-Fi networks including Michigan State University, New York University, University
of California Berkeley, and Northeastern University.

We understand that some feasibility tests and attack evaluations might be harmful to the operators and/or users. As a result, we proceed with this study in a responsible manner by running experiments in fully controlled environments. In all the experiments,
\hl{victims are always our lab members.}
Instead of aggravating the damages caused by new security vulnerabilities and attacks, 
we seek to disclose them and then provide effective solutions.

	\section{HOW DOES IT GO WRONG?}
\label{sect:how_does_it_go_wrong}



In this section, we answer the aforementioned two questions in details by disclosing our four findings with their corresponding vulnerabilities. Then these vulnerabilities are analyzed and examined in the standard, operator networks and device implementations.

\subsection{Q1: Do Wi-Fi calling devices still activate Wi-Fi calling services while associating with insecure Wi-Fi networks?}
\label{subsubsect:q1}

To answer this question, we seek to examine whether the selection policy of Wi-Fi networks on the Wi-Fi calling devices consider security impacts or not, in addition to the performance issues involving the quality of Wi-Fi link and Internet connectivity. However, it is not the case due to the discovery of three security vulnerabilities, so the answer to Q1 is yes. By exploiting these vulnerabilities, an adversary is able to intercept the packets of the Wi-Fi calling service for one victim, as well as then infer service events of his/her ongoing Wi-Fi calling call and manipulate the delivery of the service packets.

\smallskip
\textbf{Vulnerability 1 (V1).} This is a design defect in the standards~\cite{GSMA-WiFi-Calling,TS24.302}. We discover that Wi-Fi calling devices are unable to exclude an insecure Wi-Fi network from their Wi-Fi network candidates
due to the network selection mechanism of the Wi-Fi calling.


There are two Wi-Fi network selection modes: manual and automatic modes.
In the manual mode, devices maintain a prioritized list of selected Wi-Fi networks, \hl{the implementation of which is vendor-specific.}
In the automatic mode, devices select their \hl{connected} Wi-Fi networks by \hl{following the guidance from} the network infrastructure based on the ANDSF (Access Network Discovery and Selection Function)~\cite{TS24.312}. The selection is mainly based on connectivity capabilities (specified in the IEEE Standard 802.11-2012~\cite{6178212}) and radio qualities (e.g., ThreshBeaconRSSIWLANLow~\cite{TS24.302}) of available Wi-Fi networks. Both modes do not consider \hl{security factors} 
of their selected Wi-Fi networks (i.e., whether the Wi-Fi networks are vulnerable or compromised). 
\hl{As a result, Wi-Fi calling devices may select insecure Wi-Fi networks for their Wi-Fi calling services.}


\smallskip
\textbf{Vulnerability 2 (V2).} This is a common \hl{security loophole in the implementation} of Wi-Fi calling devices. We find that all of our test Wi-Fi calling devices suffer from the ARP (Address Resolution Protocol) spoofing/poisoning attack, which is \hl{usually} the prerequisite of various MITM (Man-In-The-Middle) attacks. 
\hl{In the attack, the adversary masquerades as the default gateway by sending fabricated ARP reply messages which associate its MAC address } with the gateway's IP address onto a local area network. All of the local devices' packets can be then sent to the adversary instead of the gateway. 
\hl{It allows the adversary to intercept all} the packets belonging to the Wi-Fi calling devices.

\smallskip
\textbf{Vulnerability 3 (V3).} This is an operational issue from Wi-Fi calling service operators. We discover that the Wi-Fi calling is the only service carried by the IPSec channel between the UE and the ePDG (shown in Figure~\ref{fig:volte_vowifi}), and it may be exploited to leak user privacy by monitoring the channel.  
\hl{It allows the adversary to launch a side-channel attack by inferring the Wi-Fi calling service events (e.g., call status and text messaging status).}
This vulnerability exists in all of our test operators.



\subsubsection{Validation.} We next validate these three vulnerabilities.

\textbf{V1.} We validate V1 by deploying an insecure Wi-Fi network using a Wi-Fi router which \hl{does not defend against ARP spoofing attacks}.
We test whether \hl{each of our Wi-Fi calling devices} keeps connecting to it, 
while using one host to launch an ARP spoofing attack where the host masqueraded as the tested device.
Our result shows that all the tested devices \hl{do not disassociate from the router or terminate their Wi-Fi calling services including dialing calls and text messaging}.
\hl{It causes all the \emph{downlink} Wi-Fi packets towards the devices to be intercepted by the host so that they can be inspected, dropped, or manipulated with other actions.} 

\smallskip
\textbf{V2.} To validate V2, we examine whether the tested devices can be resistant to the ARP spoofing attack.
We employ a tool, EtterCap, to send spoofed ARP messages, which claim one \hl{host} as the default gateway, to all the Wi-Fi calling devices in the network.
\hl{It leads all the devices to send uplink packets to the host instead of the gateway.}
The host can thus intercept all the \emph{uplink} Wi-Fi calling packets.


\begin{figure}
	\includegraphics[width=0.85\columnwidth]{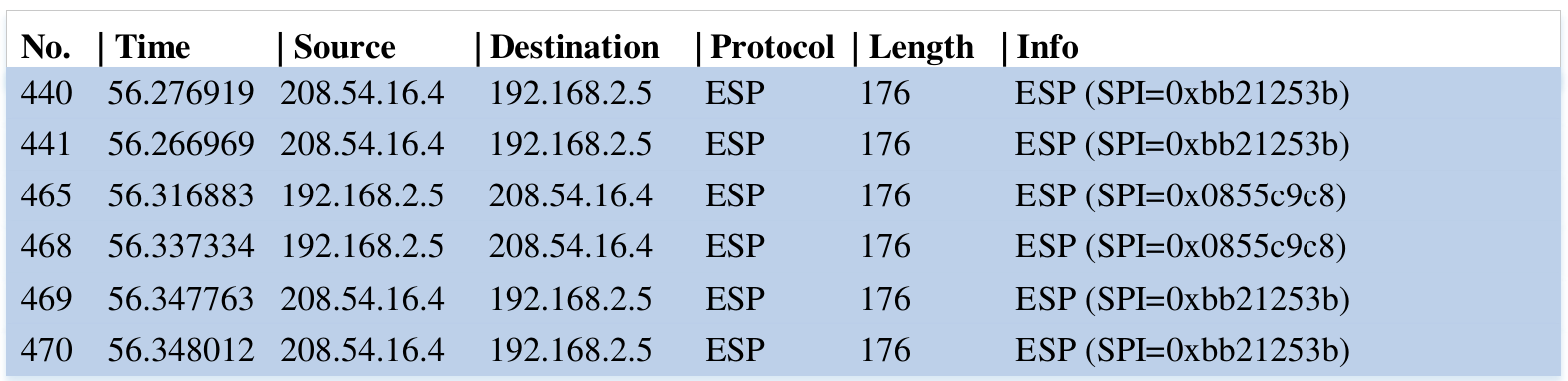}
		\caption{A trace of intercepting Wi-Fi calling packets through the ARP spoofing.}
		\label{fig:Wireshark_arp_middleman}
\end{figure}

Note that by exploiting V1 and V2, an adversary who does not have full control of Wi-Fi networks is \hl{able to intercept} all the Wi-Fi calling packets with his/her device (as shown in Figure~\ref{fig:Wireshark_arp_middleman}). However, we do not observe any alerts or warnings from the Wi-Fi calling devices. \hl{More threateningly, the adversary is not required to stay within the radio coverage (e.g., 300 feet) of the Wi-Fi APs with which victims associate.} 
More discussions will be given in Section~\ref{subsect:feas-wifi-calling-attacks}.

\smallskip
\textbf{V3.} We examine whether any information can be inferred based on the intercepted Wi-Fi calling packets, which are encrypted by the IPSec.
After analyzing their patterns, we discover that there are six service events in all the three operators' Wi-Fi calling services: dialing/receiving a call, sending/receiving a text message, and activating/deactivating the service.

\begin{figure}
	\includegraphics[width=0.8\columnwidth,height=1.7in]{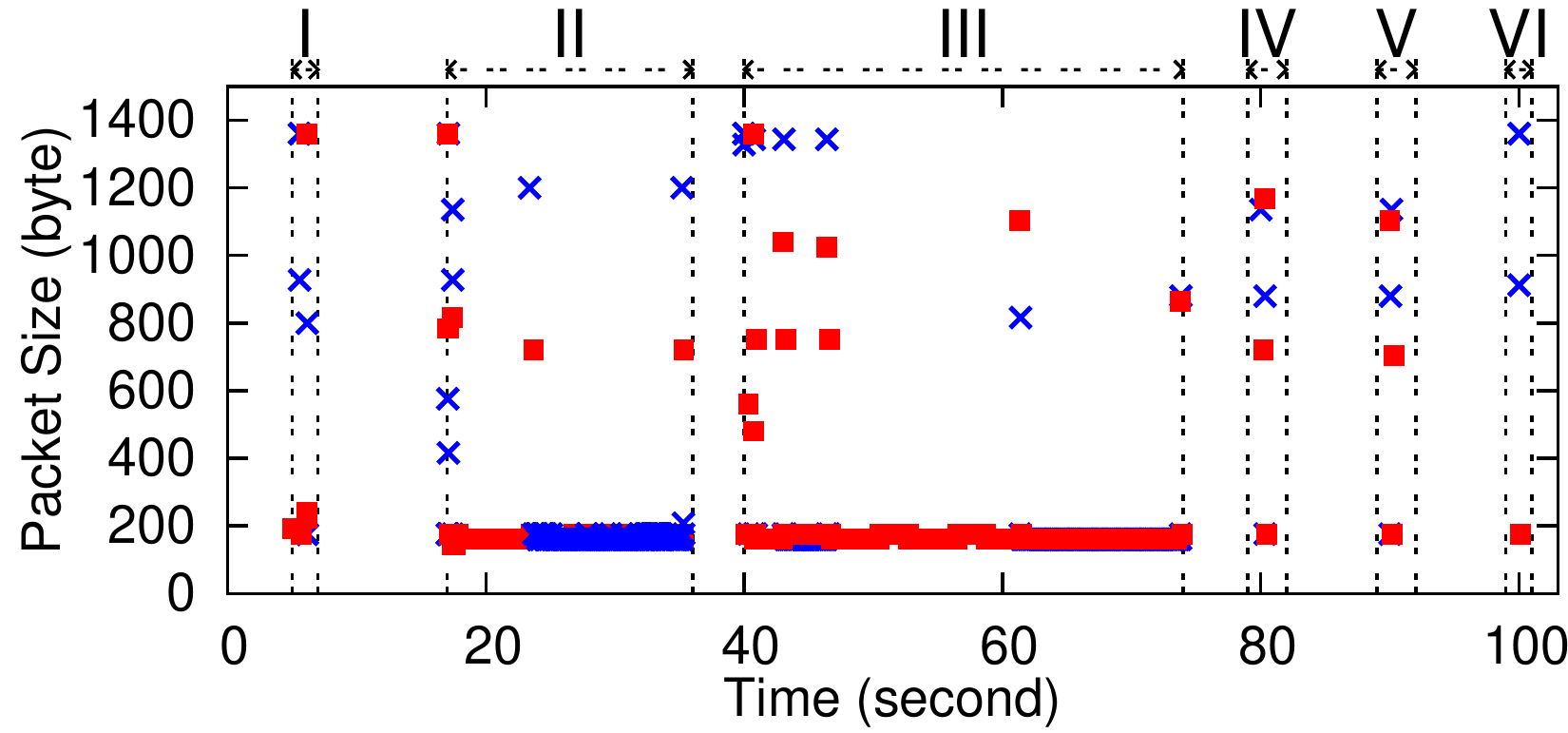}
	\caption{The IPSec packets of six Wi-Fi calling events over time ({\color{blue}{$\times$}}: uplink packets; {\color{red}{$\blacksquare$}}: downlink packets; I/VI: Activating/Deactivating Wi-Fi calling; II/III: Receiving/Dialing a call; IV/V: Sending/Receiving a text).}
	\label{fig:6events}
\end{figure}

\begin{figure}
	\includegraphics[width=0.8\columnwidth,height=1.4in]{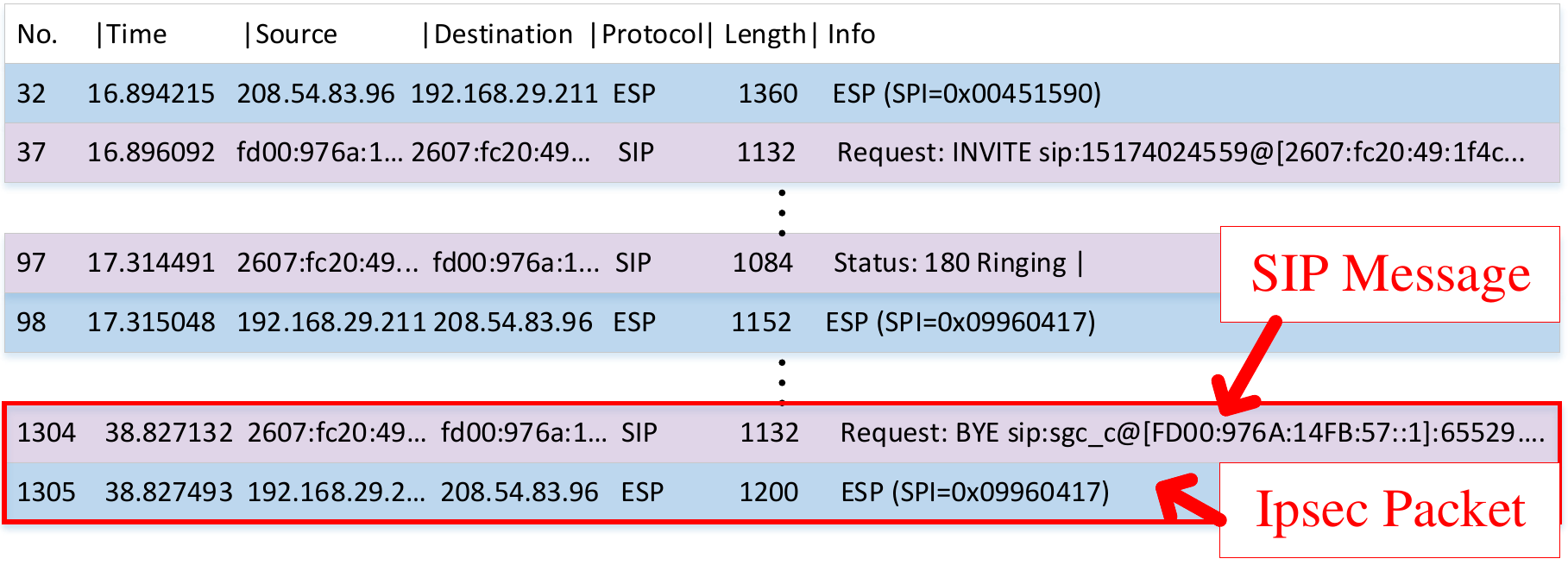}
	\caption{A trace of Wi-Fi calling packets: SIP and IPSec packets collected on a test phone.} 
	\label{fig:wireshark_capture}
\end{figure}

Figure~\ref{fig:6events} shows the IPSec packets captured on our Wi-Fi AP from an experiment, where we trigger the aforementioned six events on a test phone. It is observed that all the events differ from each other in terms of various traffic patterns, which are composed of packet direction (uplink or downlink), packet size, and packets interval.
\hl{In order to identify and classify them based on encrypted Wi-Fi traffic, we apply a decision tree method, the C4.5 algorithm~\cite{DBLP:books/mk/Quinlan93}.}
To prepare a set of training data for the C4.5 algorithm, we repeat the aforementioned six Wi-Fi calling service operations on the test phone with 50 runs and collect all the IPSec packets of the phone on the Wi-Fi AP. Based on the training data, we use the C4.5 to generate a model to \hl{classify the events}. We validate the result of the decision tree for each test by comparing it with the test phone's trace, as shown in Figure~\ref{fig:wireshark_capture}. \hl{We can get 100\% accuracy for 50 tests. Note that the trace is collected from the phone, Nexus 6P, with T-Mobile Wi-Fi calling service.}

We next examine whether \hl{the classification model works} for cross-phone/cross-carrier cases.
We \hl{consider various} devices with the Wi-Fi calling services of three carriers.
The results are summarized in Table~\ref{tab:cross_evaluation_call_event}. It is observed that those six events in all the test cases can be \hl{identified} accurately.
Specifically, the model trained based on the Nexus 6P device with T-Mobile's Wi-Fi calling service can be applied to the other devices and carriers (e.g., Samsung Galaxy J7, S6, S7, S8, Apple iPhone6, iPhone7, and iPhone8 with AT\&T or/and Verizon).

\begin{table}
	\resizebox{0.6\columnwidth}{!}{
		\scriptsize
		\begin{tabular}{c|c|c|c}
			\hline
			\textbf{Test Device} & \textbf{T-Mobile} & \textbf{AT\&T} & \textbf{Verizon}\\
			\hline
			\hline
			Samsung J7 (Verizon) & N/A & N/A & 100\% \\
			\hline
			Samsung S6 (AT\&T) & N/A & 100\% & N/A \\
			\hline
			Samsung S7 (T-Mobile) & 100\% & N/A & N/A \\
			\hline
			Samsung S8 (AT\&T) & N/A & 100\% & N/A \\
			\hline
			Nexus 6P & 100\% & N/A & N/A \\
			\hline
			iPhone 6 & 100\% & 100\% & 100\% \\
			\hline
			iPhone 7 & 100\% & 100\% & 100\% \\
			\hline
			iPhone 8 & 100\% & 100\% & 100\% \\
			\hline
		\end{tabular}
	}
	\vspace{0.2cm}
	\caption{Classification accuracy of Wi-Fi calling events in various cross-phone/cross-carrier cases. N/A means that the test phone does not support the carrier's Wi-Fi calling service.}
	\label{tab:cross_evaluation_call_event}
\end{table}

\subsubsection{Rationale and security implications}
It is not without reasons that the Wi-Fi calling standards consider only the quality of Wi-Fi links and Internet connectivity for the selection of Wi-Fi networks, since Wi-Fi calling sessions have been protected by IPSec with end-to-end confidentiality and integrity protection. Though it is unlikely for an adversary to decrypt/alter the Wi-Fi calling packets, intercepting those packets for the further attacks are still possible.
We believe that 3GPP and GSMA shall revisit current Wi-Fi network selection mechanism for the Wi-Fi calling service in terms of security and make further revisions to address the issues. Otherwise, the Wi-Fi calling users are being exposed to potential security threats.

\subsection{Q2: Do any security mechanisms exist on the devices or/and the network infrastructure to defend against \hl{possible} security threats?}
\label{subsubsect:q2}

The answer is yes. There exists a system-switch mechanism in which when a Wi-Fi calling device cannot be reached through its connected Wi-Fi network (e.g., the Wi-Fi calling signaling packets cannot be delivered to users successfully), it would switch back to the cellular network and use the cellular-based voice service (e.g., VoLTE (Voice over LTE)). This mechanism can protect the Wi-Fi calling device from the DoS attack where the Wi-Fi calling packets are discarded, because the attack impact will be considered as the case that the device is unreachable. However, they do not completely eliminate the security threats. It does not work for some attack cases, e.g., packets are dropped during an ongoing Wi-Fi calling service. The vulnerability we discovered is as follows.

\smallskip
\textbf{Vulnerability 4 (V4).} This is a design defect in the Wi-Fi calling standards. The Wi-Fi calling standards and
cellular network standards~\cite{TS23.216,TS23.237,TS24.237} indeed stipulate the procedures which provide users with the seamless service
continuity across the Wi-Fi calling and the cellular-based voice services. However, the initiation of the procedures
only considers the quality of Wi-Fi links (Vulnerability 4 (V4)). That is, once the link quality is good, the Wi-Fi calling device will not switch from the Wi-Fi calling service to the cellular-based voice even if all the Wi-Fi calling packets keep being dropped. As a result, an adversary is able to get Wi-Fi calling users stuck in malicious Wi-Fi networks and cause them to suffer from poor voice services.

\subsubsection{Validation}
We conduct experiments to validate V4 by examining the behaviors of the service continuity mechanism during DoS attacks.
We discard a device's packets during two different times of its Wi-Fi calling call:
(1) the time that a call is being dialed, and (2) the time that a call is ongoing.
Note that in our experiments, the Wi-Fi signal strengths at the Wi-Fi calling devices are strong.

\begin{figure}
	\includegraphics[width=0.8\columnwidth,height=1.3in]{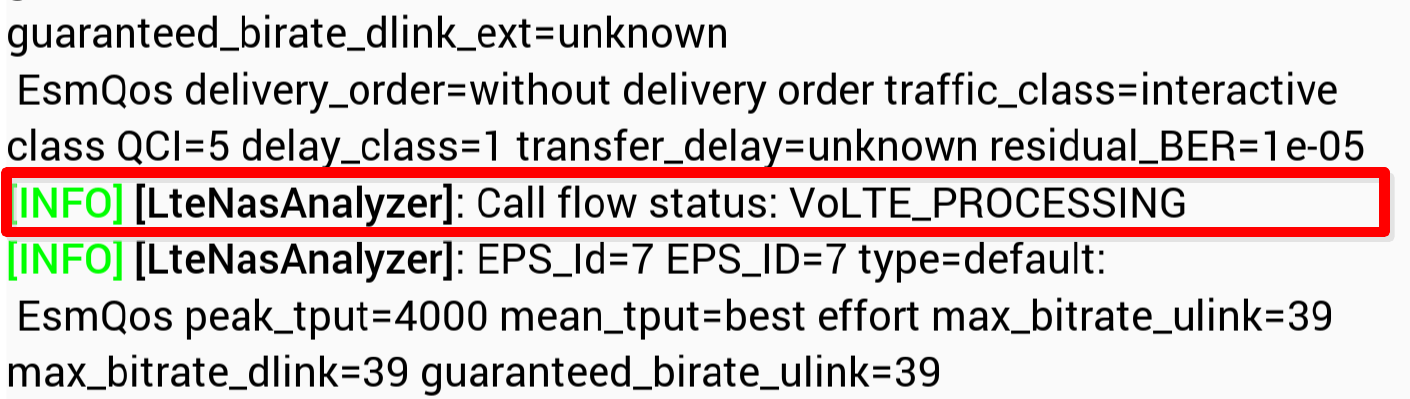}
	\caption{A trace shows that a Wi-Fi calling phone initiates a VoLTE call under the DoS attack which is launched during the time that a call is being dialed..}
	\label{fig:mobileinsight_volte_process}
\end{figure}

In the first case, the device keeps sending \texttt{SIP INVITE} packets to the Wi-Fi calling server and waiting for
the response (i.e., \texttt{SIP 100 Trying}). Since the packets are dropped, it does not receive any response to \texttt{SIP INVITE}. After six attempts, it switches back to the cellular-based voice service by initiating a VoLTE outgoing call.
{In 10 runs, all the calls are switched to VoLTE successfully, as shown in Figure~\ref{fig:mobileinsight_volte_process}. It is a low-level cellular network trace that shows the switch back to the VoLTE call. Note that it is obtained on our test phone via the MobileInsight~\cite{mobileinsight} tool, which is a tool collecting cellular network protocol traces.}

In the second case, Wi-Fi calling voice calls are interrupted within 8-10 seconds after the packets drop starts, but do not switch back to VoLTE calls. 
It shows that the service continuity mechanism does not support ongoing Wi-Fi calling calls while Wi-Fi signals are strong. 

\subsubsection{Rationale and security implications.}

Seemingly, it is an operational slip of operators, since cellular network standards have stipulated how to keep service continuity across different radio access technologies (e.g., Wi-Fi and LTE). There are two proposed methods: SRVCC (Single Radio Voice Call Continuity)~\cite{TS23.216} and DRVCC (Dual Radio Voice Call Continuity)~\cite{TS23.237,TS24.237}. However, after a second thought, it might not be the case. The SRVCC/DRVCC procedure is initiated by the network infrastructure and triggered based on the radio quality conditions of current serving base station (BS) and neighboring BSes.
It implies that the SRVCC/DRVCC will be never triggered if the radio quality of the current serving BS is good. This design makes sense in the traditional cellular network, but 
does not work for Wi-Fi calling services, whose traffic has to traverse the Internet, since the procedure may be needed for security issues regardless of radio quality. 
It allows an adversary to launch the man-in-the-middle attacks.
As a result, the design of service continuity should consider not only radio quality but also service quality due to the security concern.

	\section{User Privacy Leakage Attack}
\label{sect:privacy_leakage}

\hl{In this section, we devise a proof-of-concept attack that can leak the privacy of Wi-Fi calling users. 
We exploit the discovered vulnerabilities to collect call statistics (e.g., call duration, number of dialing calls, etc.) for each device or IP; in the meantime, cameras are used to identify nearby persons' behaviors related to phone usages. By considering two information sources together, a device's call statistics can be associated with a person's behaviors, e.g., a device with 5-second call duration and a person who holds his/her phone and speaks for 5 seconds. Based on such associations, the adversary can know the IP address of a specific Wi-Fi calling user, and further be able to inspect his/her packets. The packet inspection allows the adversary to infer user privacy including device activities (e.g., accessing gmail), device information (e.g., iPhone 7), and running applications (e.g., WeChat), etc. In addition, several prior arts have demonstrated that the call statistics can be exploited to infer other user privacy information including mood (e.g., stressful~\cite{thomee2011mobile}), personality (e.g., conscientiousness~\cite{de2013predicting}), malicious behaviors (e.g., dialing spamming calls)~\cite{DBLP:conf/ceas/BalasubramaniyanAP07}, to name a few. As a result, this attack can cause Wi-Fi calling users to leak the call statistics-based and IP-based user privacy.}

\begin{figure}[t]
	\includegraphics[width=0.95\columnwidth]{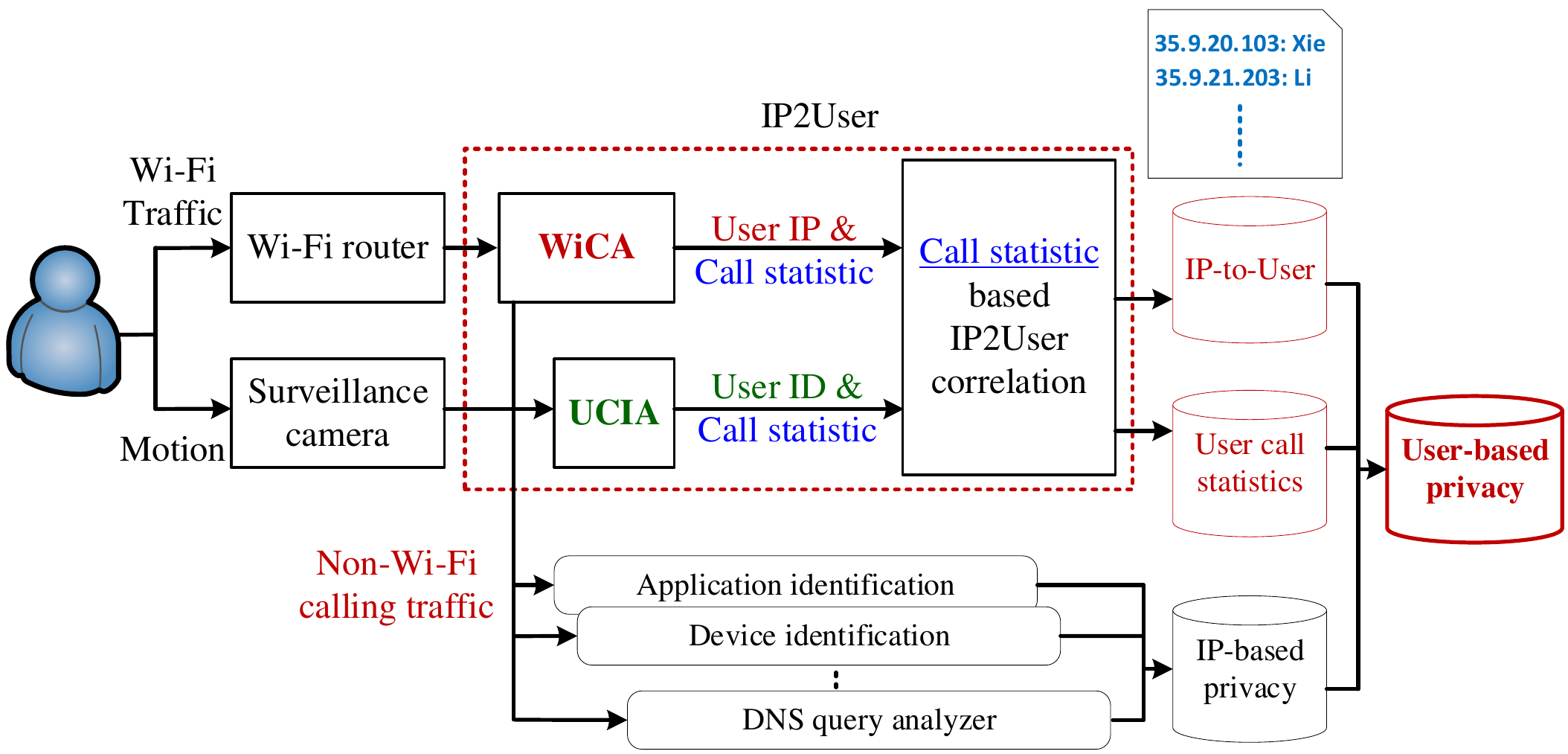}
	\caption{A Wi-Fi calling-based user privacy inference system (UPIS).}
	\label{fig:wi-fi-user-privacy-inference-sys}
\end{figure}

\begin{figure}
	\subfigure[Phone model: iPhone OS 11.1]{
		\centering
		\includegraphics[width=0.45\columnwidth, height=1in]{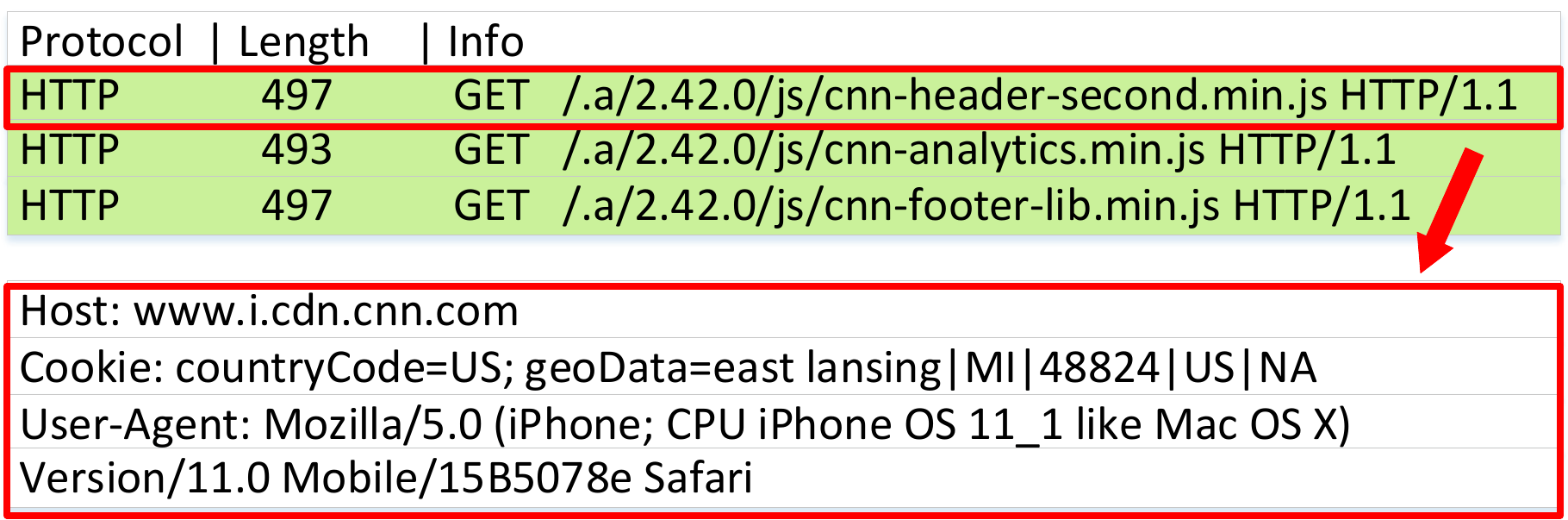}
		\label{fig:http_cnn}}
	\subfigure[Application: WeChat]{
		\centering
		\includegraphics[width=0.45\columnwidth, height=1in]{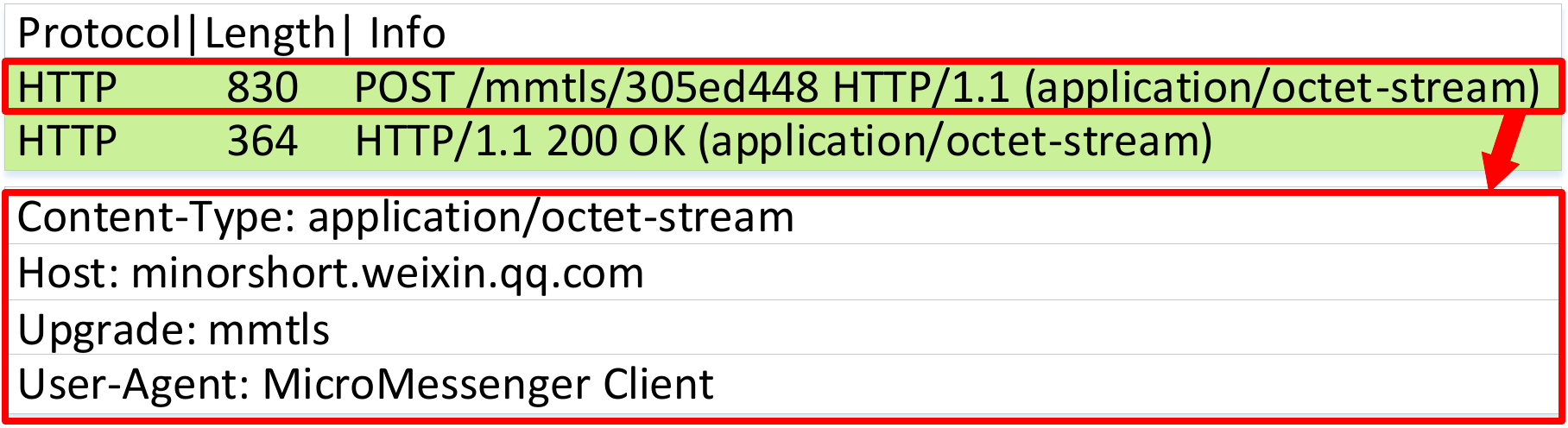}
		\label{fig:http_wechat}
	}
	\vspace{-0.2cm}
	\caption{UPIS infers phone models and running applications for a specific Wi-Fi calling user.}
	\label{fig:application_http}
\end{figure}

\hl{We develop a Wi-Fi calling-based user privacy inference system (UPIS) to launch this attack. Its system architecture is shown in Figure~\ref{fig:wi-fi-user-privacy-inference-sys}.} It consists of three major components: WiCA (Wi-Fi Calling Analyzer), UCIA (User Call and ID Analyzer), and CS-IP2U (call statistics based IP-to-User correlation) modules. \hl{WiCA intercepts all the Wi-Fi packets and classifies them based on whether they belong to the Wi-Fi calling service. For the Wi-Fi calling packets, WiCA extracts call statistics for each device IP, which includes (1) who initiates a call, (2) who hangs up first, (3) ringing time, and (4) call conversation time.
The others are dispatched to real-time traffic analyzers, which extract and collect application identification, device information, and DNS query information, etc., as shown in Figure~\ref{fig:application_http}.}
\hl{UCIA identifies users and extracts their call statistics based on face recognition and human motion detection techniques.
CS-IP2U relies on the call statistics from WiCA and UCIA to associate each user with an IP address.
It generates a mapping table with IPs and users, as well as each user's call statistics.}

\hl{We next elaborate on the WiCA, UCIA, and CS-IP2U components, and finally evaluate the UPIS system.}

\subsection{WiCA (Wi-Fi Calling Analyzer)}
\label{sect:module_wica}
\begin{figure}[t]
		\centering
		\includegraphics[width=0.95\columnwidth]{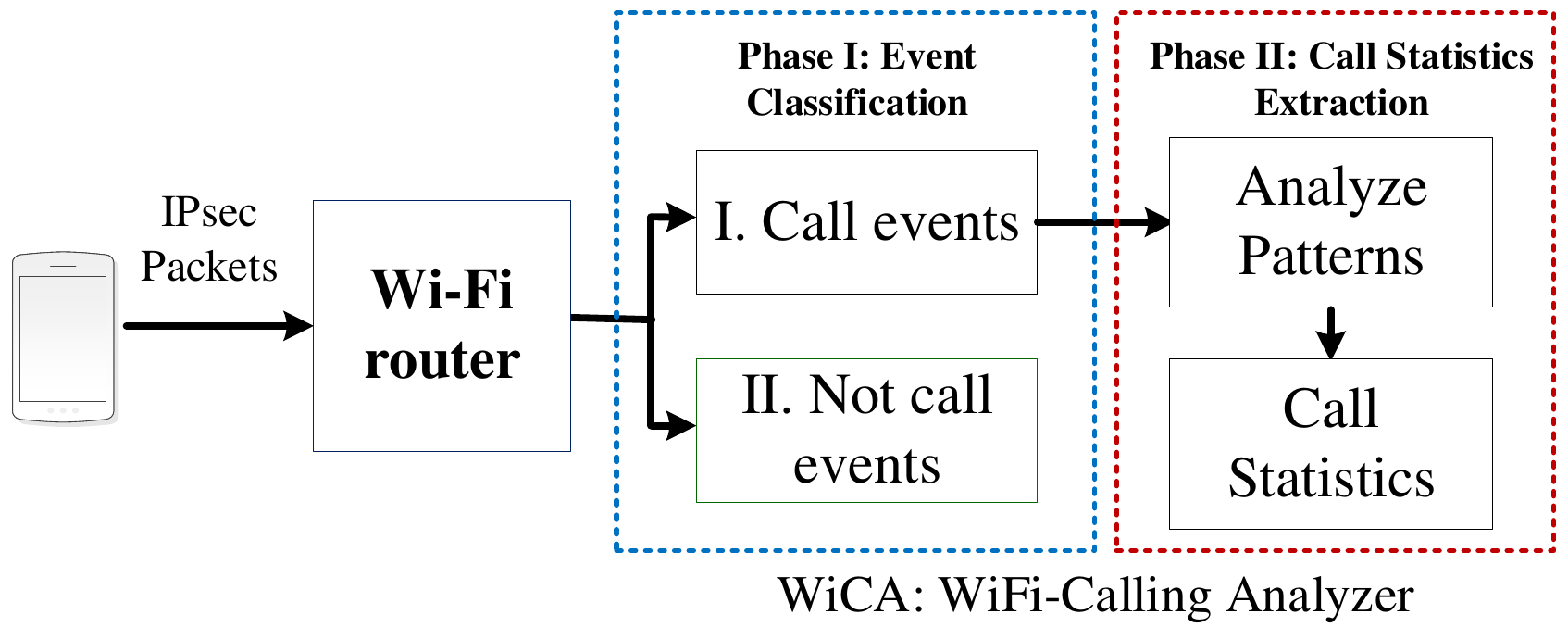}
		\caption{WiCA working flowchart.}
		\label{fig:wica_model}
\end{figure}

\hl{WiCA infers call statistics per IP basis by analyzing Wi-Fi calling traffic.
As shown in Figure~\ref{fig:wica_model}, its operation includes two working phases: (1) phase \Rmnum{1}, event classification and (2) Phase \Rmnum{2}, call statistics extraction.}
\hl{Before entering these two phases, WiCA conducts a quick analysis on the intercepted Wi-Fi packets. It identifies whether they belong to the Wi-Fi calling service by checking whether they are IPSec packets sent to/from Wi-Fi calling operators' gateways. If yes, WiCA enters those two-phase operation; otherwise, they are dispatched to other IP-based privacy analyzers.}
We next introduce each working phase.

\subsubsection{Phase I: Event Classification}

\begin{figure}[t]
	\includegraphics[width=0.7\columnwidth]{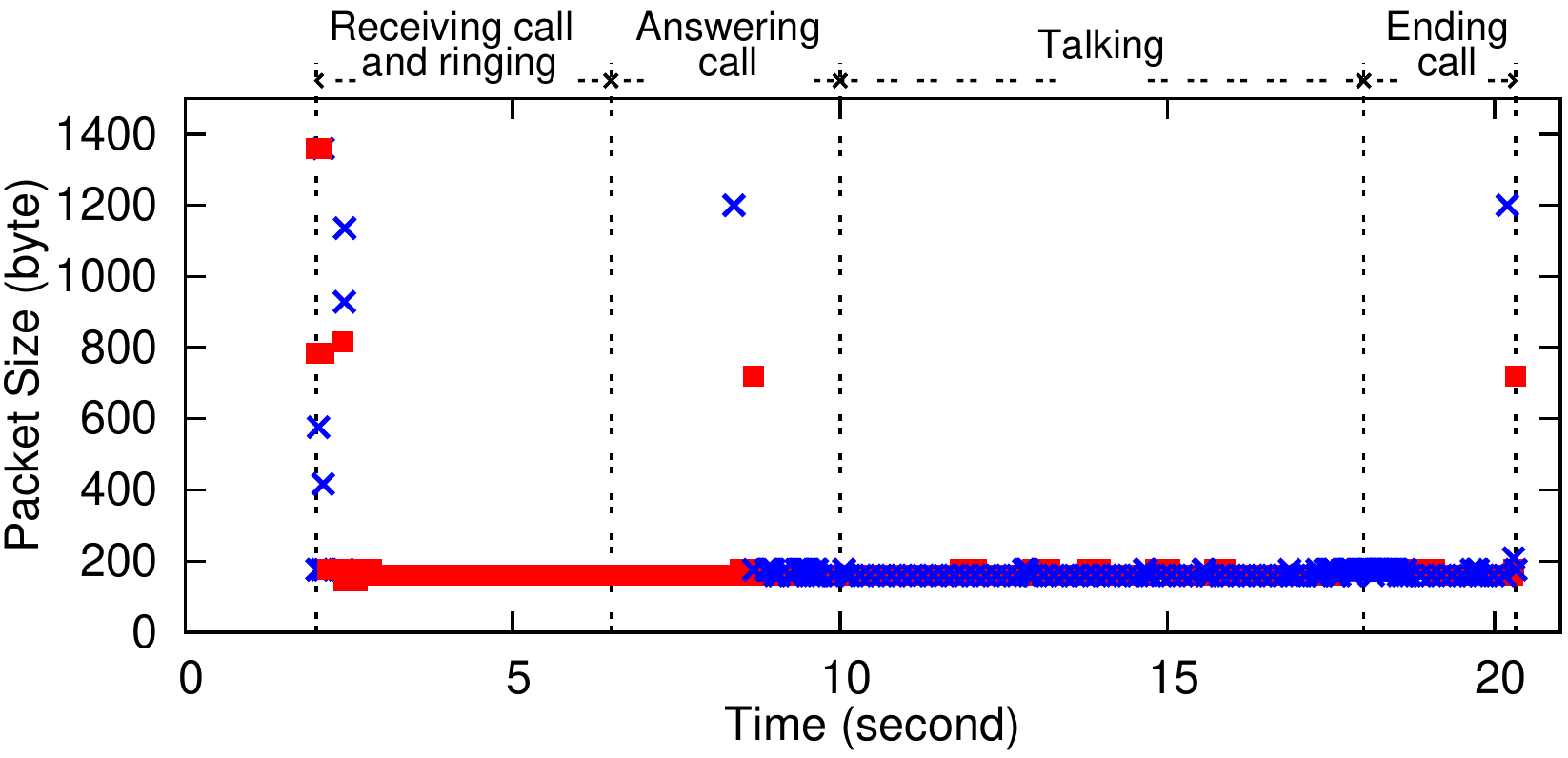}
	\vspace{-0.2cm}
	\caption{The IPSec packets observed for the callee of a call. ({\color{blue}{$\times$}}: uplink packets; {\color{red}{$\blacksquare$}}: downlink packets).}
	\label{fig:overall-callin}
\end{figure}

\hl{We identify call events based on traffic characteristics of Wi-Fi calling signaling and voice packets.
We illustrate them using an example that a user receives an incoming Wi-Fi calling call, answers it within 6 seconds after a ring tone, and then starts a 12-second voice conversation.
Figure~\ref{fig:overall-callin} shows the IPSec packets observed on the Wi-Fi AP with which the user or the callee associates. Four events can be observed:
(1) receiving a call with a ring tone; (2) answering a call; (3) talking; (4) hanging up a call.}

\begin{figure}[!h]
	\begin{minipage}[t]{0.49\linewidth}
		\subfigure[Downlink (sent by server)]{
		\centering
		\includegraphics[width=.47\textwidth]{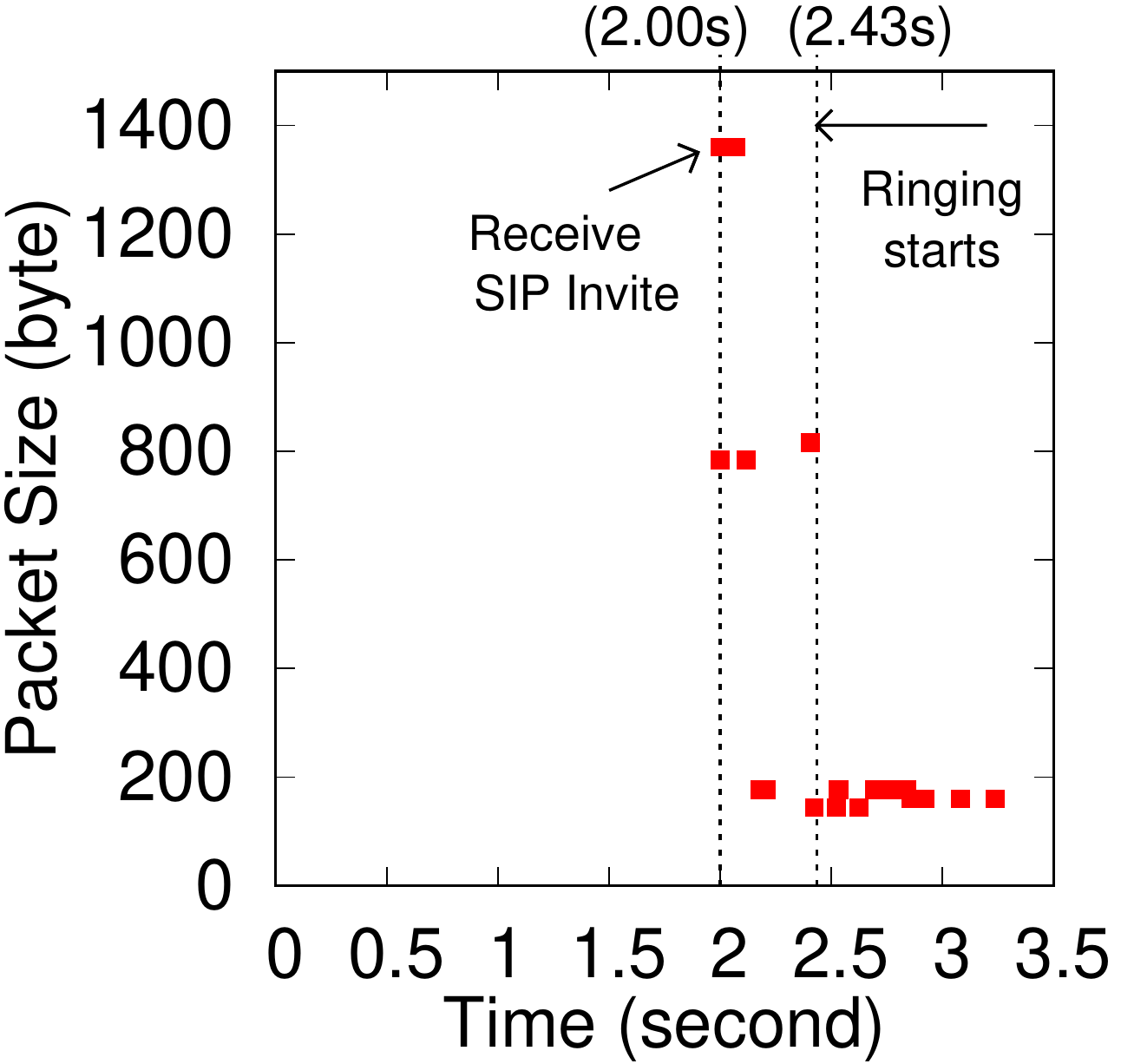}
		\label{fig:callin_receive_part1}}
		\subfigure[Uplink (sent by callee)]{
		\centering
		\includegraphics[width=0.47\textwidth]{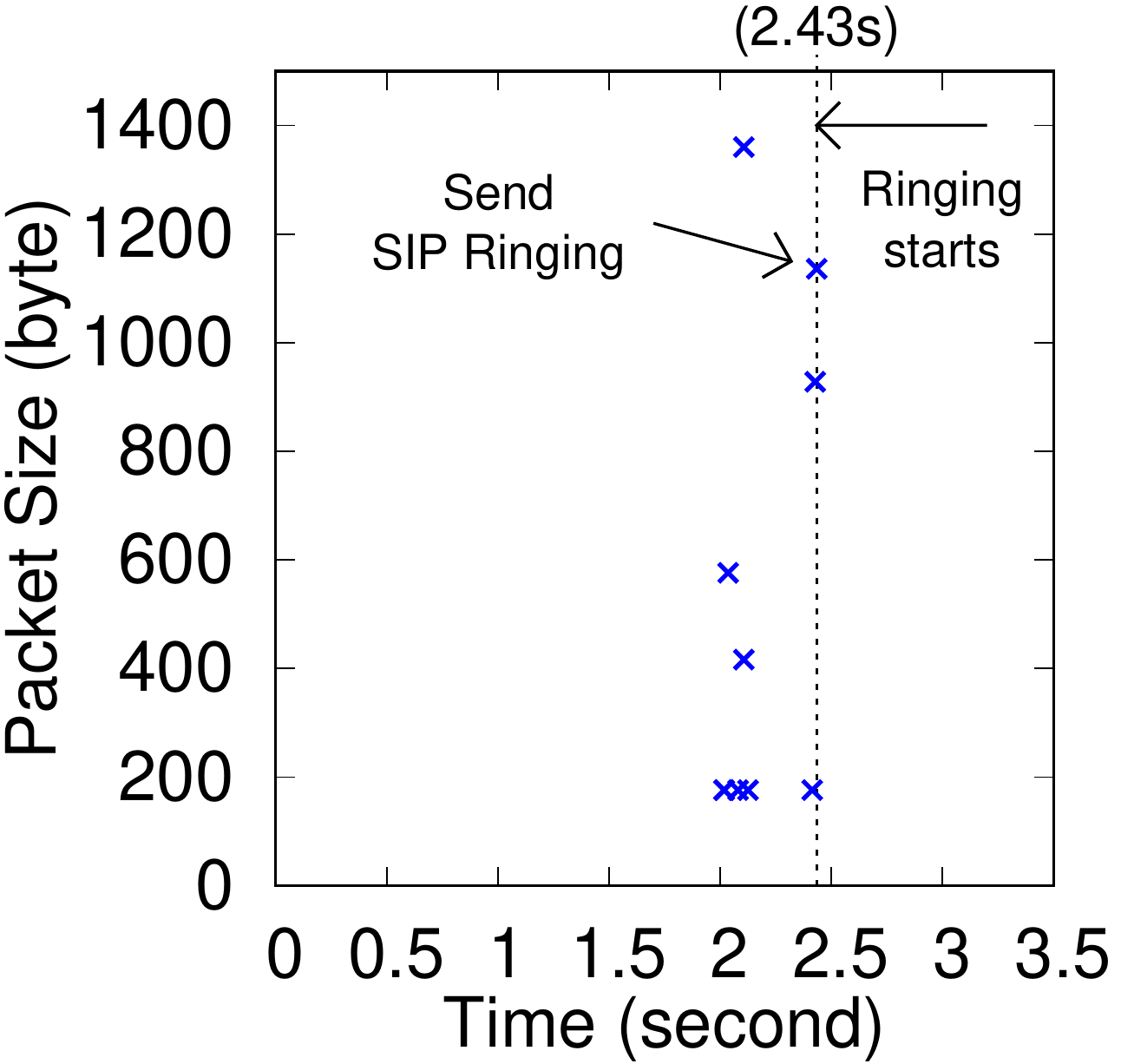}
		\label{fig:callin_send_part1}
		}
		\vspace{-0.2cm}
		\caption{Packet arrivals for the event `receiving a call with a ringtone'.}
		\label{fig:uplink-downlink-part1}
	\end{minipage}
	\hspace{0.1cm}
	\begin{minipage}[t]{0.49\linewidth}
		\subfigure[Downlink (sent by server)]{
		\includegraphics[width=0.47\textwidth]{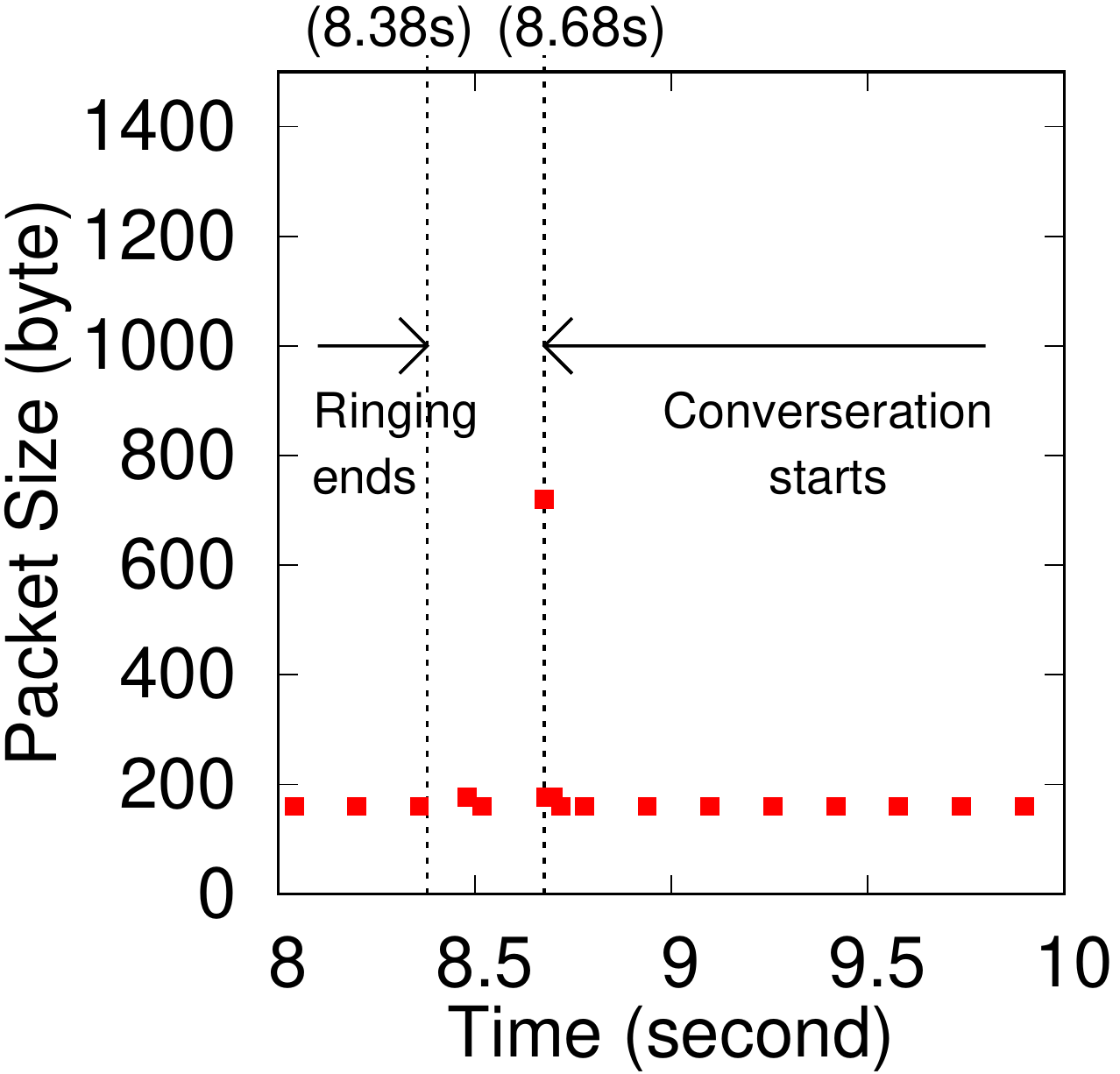}
		\label{fig:callin_receive_part2}}
		\subfigure[Uplink (sent by user)]{
		\includegraphics[width=0.47\textwidth]{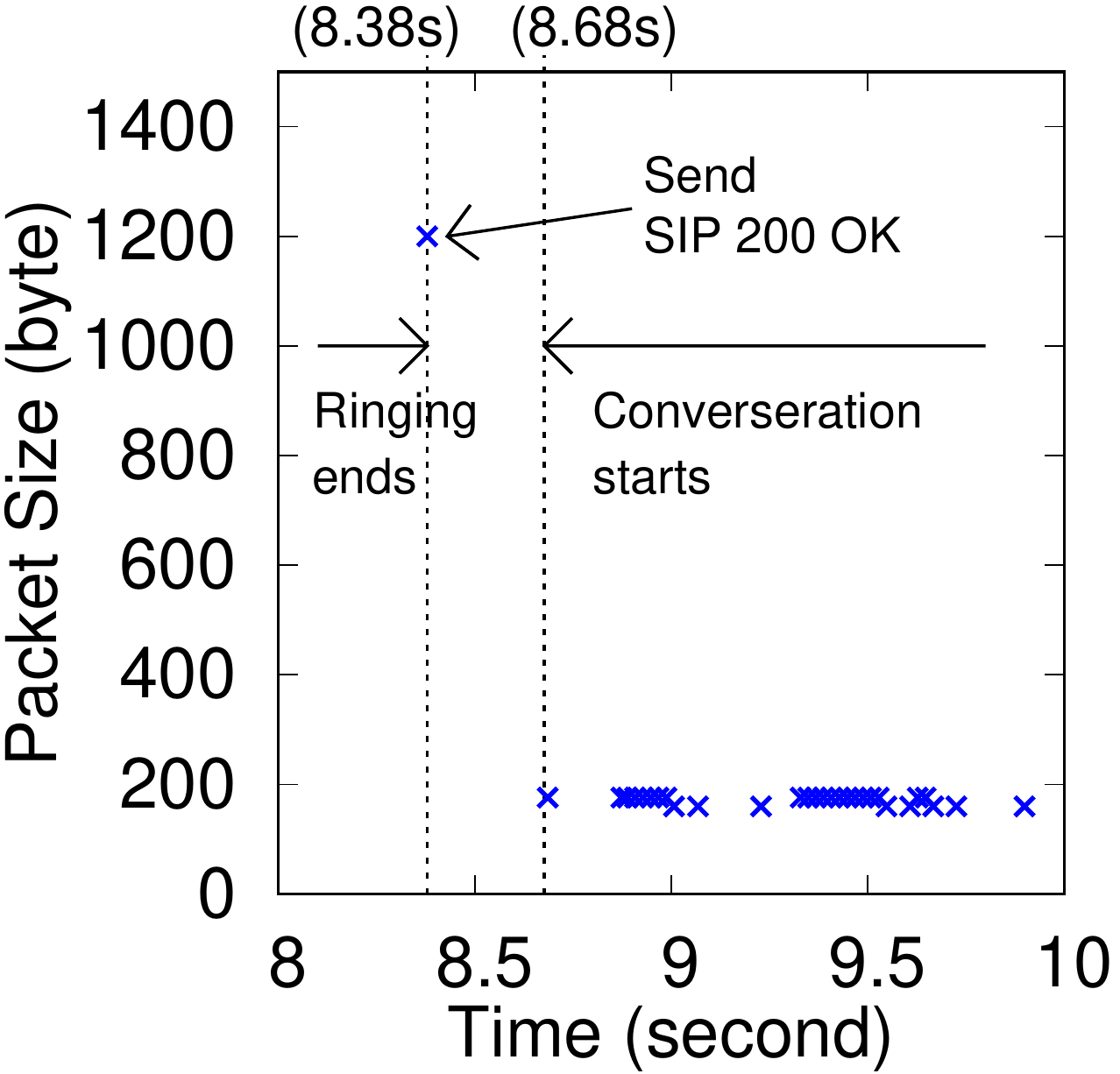}
		\label{fig:callin_send_part2}
		}
		\vspace{-0.2cm}
		\caption{Packet arrivals for `answering a call'.}
		\label{fig:uplink-downlink-part2}
	\end{minipage}
\end{figure}

\emph{Event 1: Receiving a call with a ringtone}. \hl{Figures~\ref{fig:callin_receive_part1} and ~\ref{fig:callin_send_part1} respectively show downlink and uplink packets between the Wi-Fi calling server and the callee for this event.
At the 2nd second, a 1360-byte IPSec packet is received by the callee. After decrypting it at the callee side, 
it is an SIP \texttt{INVITE} message, which indicates that a call is coming. 
At the 2.43th second, the callee sends an SIP \texttt{180 RINGING} message to the Wi-Fi calling server (Figure~\ref{fig:callin_send_part1}). Afterwards, it is observed that several small IPSec packets with only 176 bytes are received by the callee, but the callee does not send any packets back.
We discover that they are voice packets over the RTP (Real-Time Protocol) protocol.}

\smallskip
\emph{Event 2: Answering a call}
\hl{As shown in Figures~\ref{fig:callin_receive_part2} and \ref{fig:callin_send_part2},
the callee answers the call at the 8.38th second by sending an SIP \texttt{200 OK} message to the server, and then receives an acknowledgment at the 8.68th second. 
Afterwards, the call conversation starts and the callee begins to send/receive voice packets.}

\begin{figure}[!h]
	\begin{minipage}[t]{0.49\linewidth}
		\subfigure[Downlink (sent by server)]{
			\centering
			\includegraphics[width=.47\textwidth]{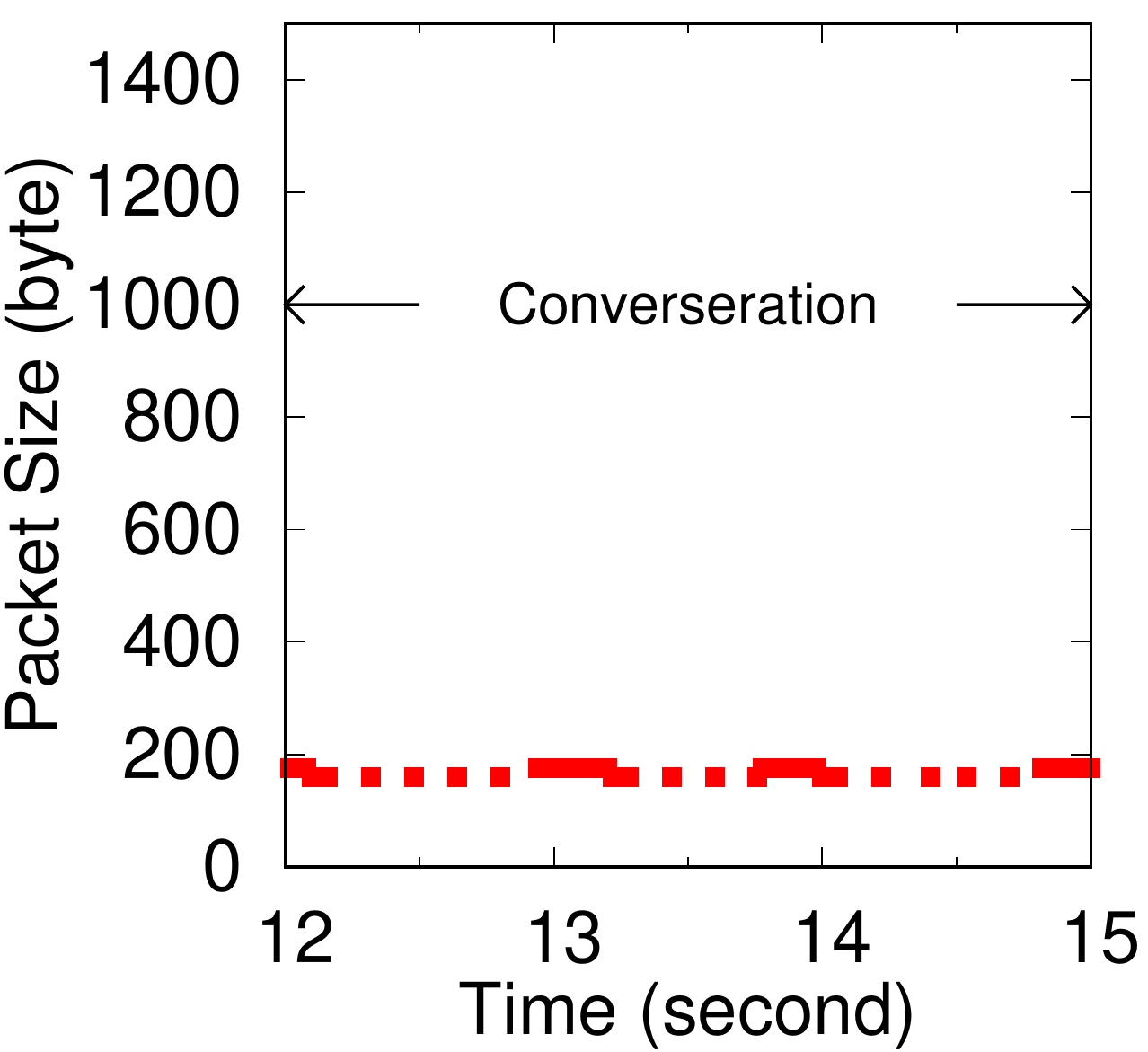}
			\label{fig:callin_receive_part3}}
		\subfigure[Uplink (sent by callee)]{
			\centering
			\includegraphics[width=0.47\textwidth]{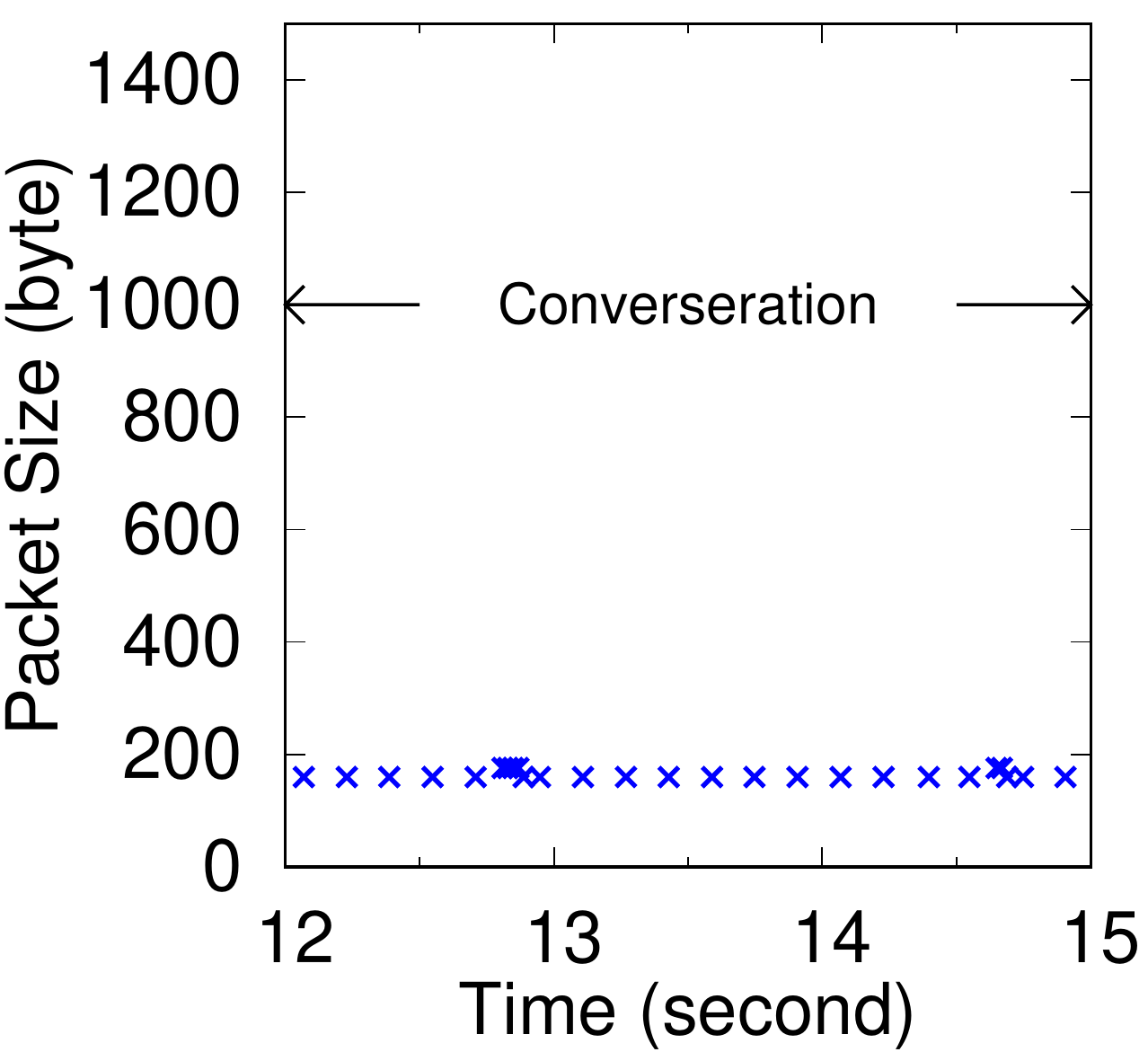}
			\label{fig:callin_send_part3}
		}
		\vspace{-0.2cm}
		\caption{Packet arrivals for `talking'.}
		\label{fig:uplink-downlink-part3}
	\end{minipage}
	\hspace{0.1cm}
	\begin{minipage}[t]{0.49\linewidth}
		\subfigure[Downlink (sent by server)]{
			\includegraphics[width=0.47\textwidth]{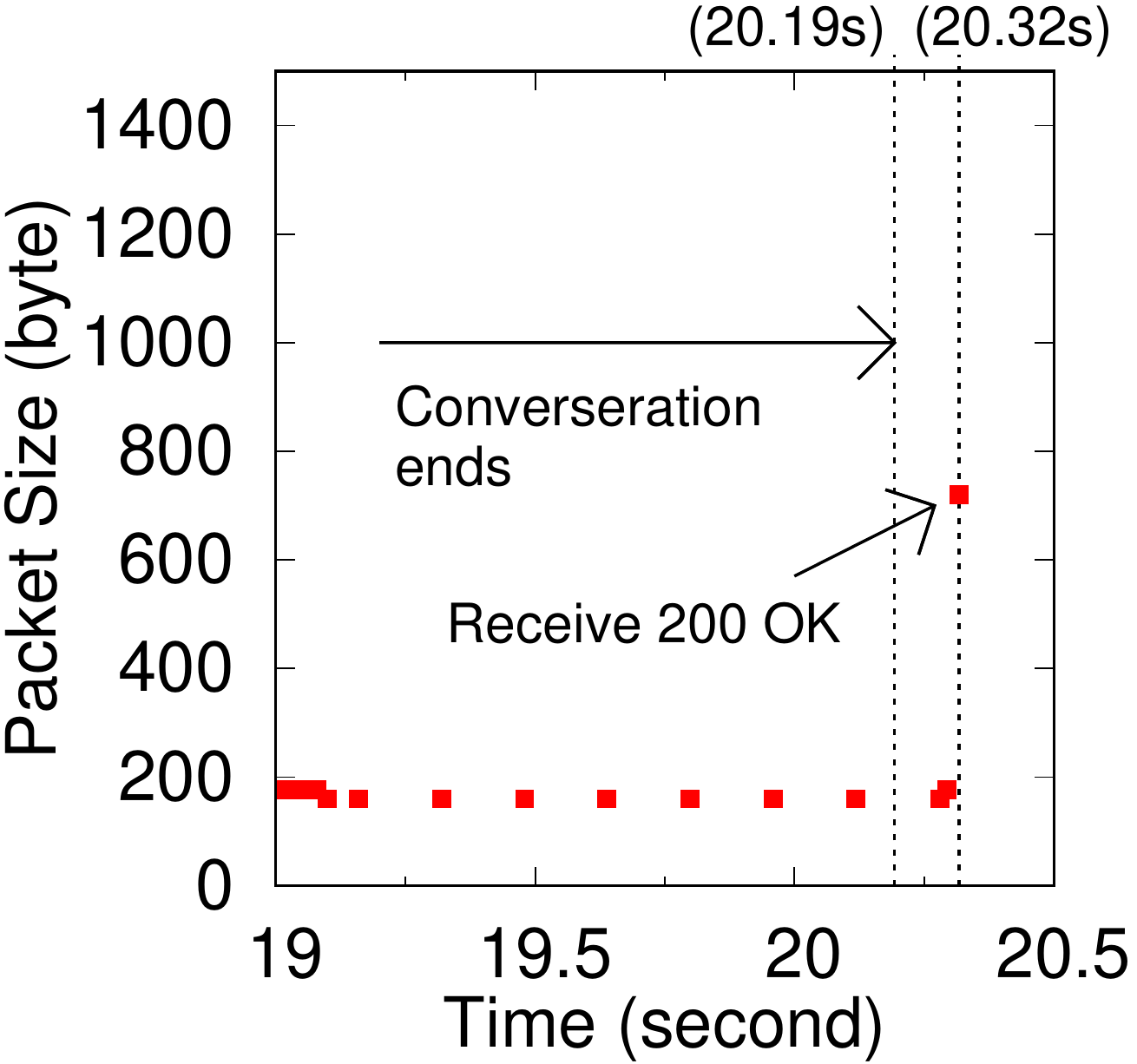}
			\label{fig:callin_receive_part4}}
		\subfigure[Uplink (sent by user)]{
			\includegraphics[width=0.47\textwidth]{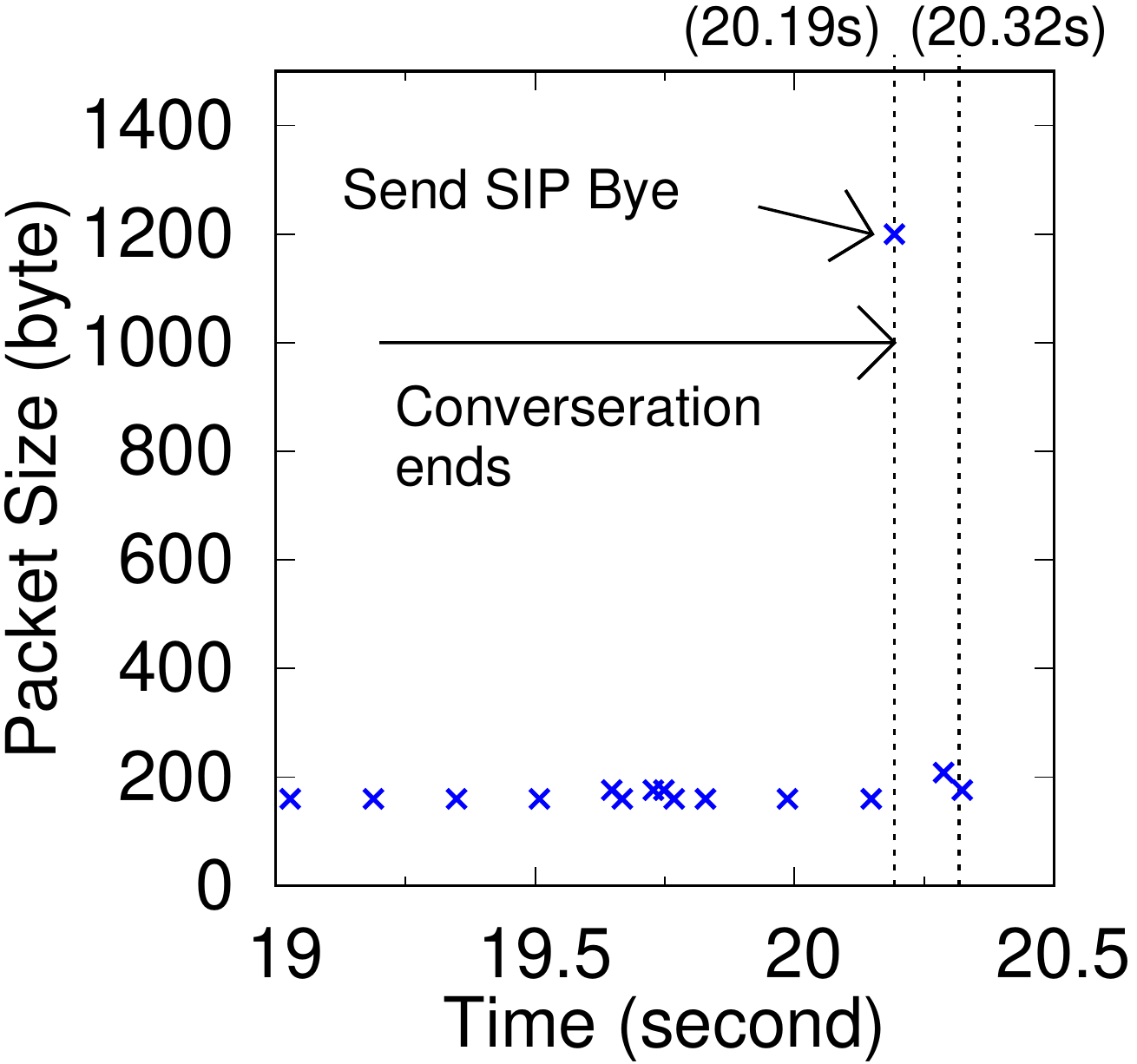}
			\label{fig:callin_send_part4}
		}
		\vspace{-0.2cm}
		\caption{Packet arrivals for `hanging up a call'.}
		\label{fig:uplink-downlink-part4}
	\end{minipage}
\end{figure}

\smallskip
\emph{Event 3: Talking} \hl{The observed traffic for this event is shown in Figure~\ref{fig:uplink-downlink-part3}. During the call conversation, the callee keeps sending/receiving voice packets to/from the Wi-Fi calling server, and no SIP message 
is observed. We further discover that the callee at least receives 10 voice packets every two seconds from the server.}

\smallskip
\emph{Event 4: Hanging up a call} 
\hl{The callee sends a \texttt{BYE} message at the 20.19th second when hanging up the call, as shown in Figure~\ref{fig:callin_send_part4}.
After the 20.32nd second, no more IPSec packets are observed. Note that when the \texttt{BYE} is sent by the server, it means that the caller hangs up first.} 


\hl{\textbf{Traffic Pattern Analysis.} We have five observations on the traffic characteristics of Wi-Fi calling voice calls.}
\begin{enumerate}
	\item The sizes of IPSec packets carrying voice are smaller than 200 bytes (e.g., 176 bytes).
	\item The sizes of IPSec packets, which carry critical Wi-Fi calling signaling messages (i.e., \texttt{INVITE}, \texttt{180 RINGING}, \texttt{200 OK}, \texttt{BYE}), are much larger than voice packets (e.g., 800-1360 bytes).
	\item The callee receives voice packets from the Wi-Fi calling server after the \texttt{180 RINGING} message is sent.
	\item No voice packets are sent out by the callee before the call conversation starts.
	\item The callee keeps receiving more than 10 voice packets every two seconds from the Wi-Fi calling server after the call conversation starts.
\end{enumerate}

Based on the analysis results, WiCA can identify whether an outgoing call is initiated, an incoming call arrives or a call ends, and further classify the IPSec packets into the \texttt{Call events} and \texttt{Not call events} categories. Note that these five observations except the third one are \hl{confirmed to be} available for all our test operators. The third observation is only applicable to T-Mobile and AT\&T. 

\subsubsection{Phase II: Call Statistics Extraction}
\label{subsubsect:phase2}
\begin{figure}[h!]
	\centering
	\includegraphics[width=0.55\columnwidth, height=0.7in]{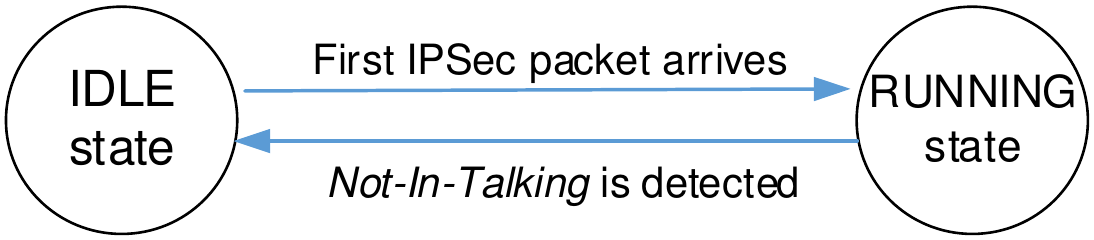}
	\caption{State transition diagram of WiCA.}
	\label{fig:pa-fsm}
\end{figure}

In Phase II, WiCA seeks to extract call statistics from IPSec packets. Figure~\ref{fig:pa-fsm} illustrates its finite state machine, where the initial state is \texttt{IDLE}. It works as follows.

\textbf{Step 1.} At the initial \texttt{IDLE} state, when any IPSec packet is received, the system starts to do event extraction by moving to the \texttt{RUNNING} state.
\hl{The system checks whether the IPSec packet belongs to a call event based on whether it is sent to/from the Wi-Fi calling server.
If yes, the event, `dialing a call' or `receiving a call', is classified based on its forwarding direction. In the former, the IPSec packet is sent from the UE; otherwise, it is sent to the UE.}

\textbf{Step 2.}
\hl{At the \texttt{RUNNING} state, WiCA uses a 2-second time window to group the collected IPSec packets and classifies them by the following three categories: \emph{C-Large}, \emph{C-Middle}, and \emph{C-Small}.
They represent the packets with the sizes larger than 800~bytes, between 200 and 800 bytes, and smaller than 200 bytes, respectively.
The \emph{C-Large} includes some critical SIP call messages (e.g., \texttt{INVITE}, \texttt{RINGING}, etc.), whereas the \emph{C-Small} has voice packets.
Note that the 2-second packet collection is denoted as $Data_{2sec}(x)$, where $x$ is the sequence of a series of the 2-sec collection sets.}

\textbf{Step 3.} \hl{WiCA can identify three scenarios, \emph{Ringing}, \emph{Talking} and \emph{Not in Talking}, based on the number of uplink and downlink \emph{C-Small} packets in $Data_{2sec}(x)$, which are denoted as $Num\_UL\_{C_{Small}}$ and $Num\_DL\_{C_{Small}}$ respectively.
The rules are summarized in Table~\ref{tab:rules-determines-scenarios}.
When no scenario is identified, $Data_{2sec}(x)$ is buffered and the system moves back to \texttt{Step 2}.
When one of these three scenarios is identified in the $Data_{2sec}(x)$ collection, the subsequent actions are taken in the following.}

\begin{table}
	\begin{minipage}{1\columnwidth}
		\begin{center}
			\resizebox{0.6\columnwidth}{!}{
				\begin{tabular}{c|c|c}
					\hline
					\multicolumn{2}{c|}{\textbf{Conditions}}   &  \multirow{2}{*}{\textbf{Identified Scenarios}}   \\
					\cline{1-2}
					$Num\_UL\_{C_{Small}}$ & $Num\_DL\_{C_{Small}}$ & \\
					\hline
					\hline
					=0 & $>$10 & Ringing\footnote{\hl{Only applicable to AT\&T and T-Mobile but not Verizon, which does not send small voice packets to the Wi-Fi calling callee when his/her phone is ringing.}}\\
					\hline
					$>$10 & $>$10 & Talking\\
					\hline
					=0 & =0 & Not in Talking\\
					\hline
				\end{tabular}
			}
			\vspace{0.1cm}
			\caption{$Num\_UL\_{C_{Small}}$ and $Num\_DL\_{C_{Small}}$, which respectively represent numbers of uplink and downlink packets smaller than 200 bytes within 2 seconds, are used to determine \emph{Ringing}, \emph{Talking}, \emph{Not in Talking} scenarios for Verizon, T-Mobile and AT\&T.}
			\label{tab:rules-determines-scenarios}
		\end{center}
	\end{minipage}
\end{table}

\begin{itemize}
\item \emph{Ringing:}
\hl{the system revisits the last collection, $Data_{2sec}(x-1)$, and looks for the time that the last \emph{C-Large} IPSec packet is captured. This times is considered as the time that the ring starts, $T_{RingingStart}$.} 

\item \emph{Talking:}
\hl{when no \emph{Talking} scenarios are identified before this, the system revisits the collection $Data_{2sec}(x-1)$  and finds the time that the first \emph{C-Large} IPSec packet (i.e., \texttt{SIP 200 OK}, which indicates the event of answering the call) is captured. This time is considered as the time that the talk starts, $T_{TalkingStart}$.}

\item \emph{Not In Talking:}
\hl{the system revisits the collection, $Data_{2sec}(x-1)$, to discover the time that the first \emph{C-Large} IPSec packet (i.e., \texttt{SIP BYE}) is captured. This time is considered as the time of the call end, $T_{CallEnd}$.
Moreover, if this \emph{C-Large} packet is sent by the UE, it means that the UE side hangs up first. Otherwise, it indicates that the other side hangs up first. 
When the call end is observed in the \emph{Not in Talking} event, the pattern analyzer outputs who initiates the call, ringing time duration (i.e., $T_{TalkingStart}-T_{RingingStart}$ or $T_{CallEnd}-T_{RingingStart}$), conversation time duration (i.e., $T_{CallStop}$-$T_{TalkingStart}$), and who hangs up first. The system then returns to the \texttt{IDLE} state. Note that the conversation duration estimation is not applicable to the calls where Wi-Fi calling users do not answer them or hang up them before the conversations start.}
\end{itemize}


\subsubsection{Implementation}

\hl{We implement WiCA using Python3 on a 2014 Macbook Pro laptop (CPU:Intel I5-4278U, RAM: 8GB). It consists of two modules: event classification and call statistics extraction.}

\textbf{Event Classification.} 
\hl{WiCA intercepts all the Wi-Fi packets using ARP spoofing and then filters out packets with operators' IP addresses (e.g., those of T-Mobile, AT\&T, and Verizon) and with the IPSec type. According to the above analysis, these packets belong to the Wi-Fi calling service.
WiCA buffers them into memory and classifies them into six Wi-Fi calling events based on the identified traffic patterns (i.e., packet direction, interval, packet size, etc.).
We develop this classifier based on the scikit-learn library~\cite{scikit-learn}.
The packets together with corresponding events are then delivered to the call statistics extraction module.}

\textbf{Call Statistics Extraction.} We develop a call statistics extraction module in Python3 to extract the call statistics from the packets sent by the Event Classification module. The extraction is carried out based on the traffic pattern analysis in Section~\ref{subsubsect:phase2} and Wi-Fi calling packet characteristics specified in Table~\ref{tab:rules-determines-scenarios}. It can identify who initiates a call, who hangs up a call first, and the time that  \texttt{Ringing}/\texttt{Talking} starts/stops.

\subsubsection{Evaluation} We next evaluate the WiCA's performance with three top-tier U.S. operators, Verizon, AT\&T, and T-Mobile. 
On each test phone, we dial 50 outgoing calls and receive 50 incoming calls using the Wi-Fi calling service. Meanwhile, we collect call-related traces on the phone during the experiments. 
\hl{Note that we collect the traces via a command, \texttt{logcat -b radio -v threadtime | grep "update phone state"}, on Android phones.}

We next compare the events identified by the WiCA and those extracted from the collected traces.
The results show that WiCA can correctly identify (1) who initiates the call and (2) who hangs up first in all the experiments. For the ringing and conversation time, we compare the difference between the time estimated by the WiCA and that observed on the test phones. Table~\ref{tab:cross_evaluation_call_event} summarizes the mean and standard deviation results of estimation errors. \hl{It is observed that all the errors are smaller than 0.8s in the experiments.}
\hl{Note that we are currently not able to determine the ringing time for Verizon since its network does not send small voice packets to the Wi-Fi calling callee after his/her phone rings.} 

\begin{table}
	\begin{minipage}{1\columnwidth}
		\begin{center}
			\resizebox{0.7\columnwidth}{!}{
				\scriptsize
				\begin{tabular}{c|c|c|c|c|c|c}
					\hline
					\multirow{2}{*}{\textbf{Time}} & \multicolumn{2}{c|}{\textbf{T-Mobile}} & \multicolumn{2}{c|}{\textbf{AT\&T}} & \multicolumn{2}{c}{\textbf{Verizon}}\\
					\cline{2-7}
					& Mean & Std & Mean & Std & Mean & Std \\
					\hline
					\hline
					Ringing & 0.16s  & 0.11s & 0.34s & 0.11s & N/A& N/A \footnote{Verizon does not send small voice packets to the Wi-Fi calling callee when his/her phone is ringing.}\\
					\hline
					Conversation& 0.17s  & 0.07s & 0.67s& 0.13s & 0.44s& 0.2s \\
					\hline
					\hline
				\end{tabular}
			}
			\vspace{0.1cm}
			\caption{Errors of ringing and conversation time estimation.}
			\label{tab:cross_evaluation_call_event}
		\end{center}
	\end{minipage}
\end{table}

\subsection{UCIA (User Call and ID Analyzer)}
\label{sect:module_uca}

UCIA is a visual recognition system which \hl{identifies users} and \hl{user motions related to making calls} (e.g., a user moves a phone to his/her ears). It leverages multiple computer vision technologies including a face detector designed for finding small faces in a video, called tiny face detector, DR-GAN (Disentangled Representation learning-Generative Adversarial Network), HOG (Histogram of Oriented Gradient), and SVM (Support Vector Machine). \hl{It can precisely output user identities and specific calling motions. 
Note that UCIA does not require users to still or a high-resolution video. It can support the video that contains the face resolutions higher than or equal to 25x10~\cite{hu2017finding}}.

\begin{figure}[t]
	\includegraphics[width=0.95\columnwidth]{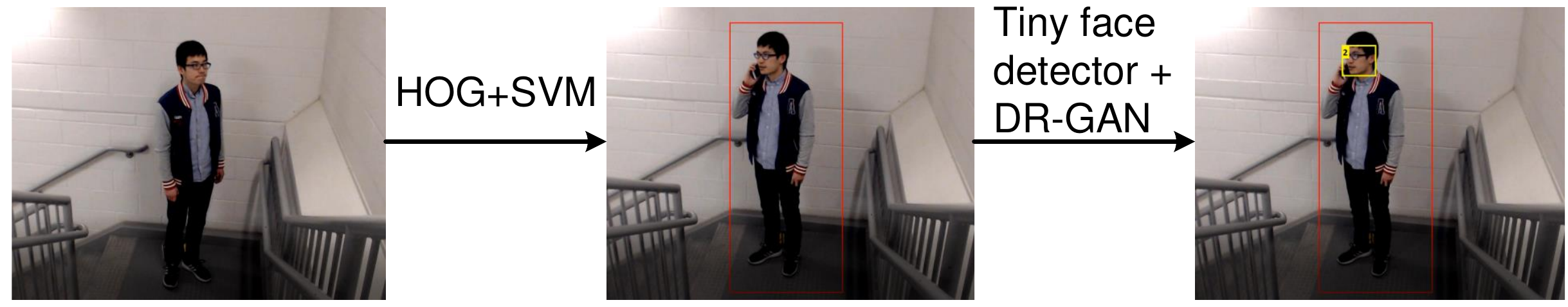}
	\caption{A working flowchart of the UCIA. The red bounding box denotes a detected user calling/talking motion, whereas the yellow bounding box denotes a detected user face.}
	\label{fig:visual_detection}
\end{figure}

Figure~\ref{fig:visual_detection} illustrates the working flow chart of the UCIA, which analyzes videos on a per-frame basis. It consists of two modules: (1) calling/talking motion detection and (2) user face detection and recognition. \hl{In each video frame, it uses the HOG and SVM models to detect calling/talking motions for all users, and labels those whose motions are detected using red bounding boxes.
For each red bounding box, it furthers uses the tiny face detector and DR-GAN model to label the user face with a yellow bounding box, and identifies his/her identities (i.e., names).}
\hl{We next detail these two modules and evaluate the UCIA's performance.}

\subsubsection{Calling/talking Motion Detection}
\label{sect:motion_recognition}
\hl{This module recognizes calling/talking motions using SVM and HOG techniques.}

\textbf{SVM.} \hl{We train an SVM model to recognize whether a person is dialing a call or talking in a call.
Since no video sets contain those two actions, we invite a number of volunteers to record videos of their dialing/talking motions.
To enable the model to differentiate those two actions from the others, we do the training by mixing those recorded videos with the UCF101 database~\cite{soomro2012ucf101} that has 13320 videos from 101 action categories.}

\textbf{HOG.}
It has been a very famous detection algorithm since 2005~\cite{dalal2005histograms}. An HOG descriptor is obtained by computing the statistic information of histograms of oriented gradient for the local areas of an image. Its applications have been widely used in object recognition, especially gaining a great success in the pedestrian detection.
Specifically, object appearances and shapes in an image can be well-represented by the density and orientation of the gradient or edge. 
To implement the HOG descriptor, we first divide the image into different small connected components, called cells, and then collect the orientation histogram of gradients for each pixel within each cell. Finally, we concatenate all the histograms to get the final HOG descriptor.

\hl{The HOG descriptor~\cite{dalal2005histograms} is used as the feature representation of persons. Each frame in a surveillance video generates a lot of candidate bounding boxes by a sliding window. After all the persons are marked by the bounding boxes, the pre-trained SVM classifier determines whether a dialing action happens in each bounding box based on the change of edge information (i.e., gradient information). As a result, the HOG descriptor is a reasonable choice to work with the SVM model to recognize the dialing/talking actions.}


\subsubsection{User Face Detection and Recognition}

\hl{We deploy a face recognition model to detect and recognize the person face. It is based on the technique of deep convolution neural network (CNN), which is getting more and more popular with its high accuracy and efficiency in handling most vision problems. Since not all of surveillance cameras offer high video quality (e.g., 1080p), we adopt a tiny face detector~\cite{hu2017finding}, which is designed to detect small faces (e.g., a face with the size of $3*3cm^2$) in a low-resolution video. Note that the tiny face detector can also support large faces in a high-resolution video.
Moreover, since people do not always face to the camera in a frontal view, extracting pose-invariant feature representations is critical to improve the face recognition performance. DR-GAN~\cite{tran2017disentangled} can generate those representations and then recognize user identities. We next introduce the tiny face detector and the DR-GAN.}



\begin{figure}[t]
	\includegraphics[width=0.95\columnwidth]{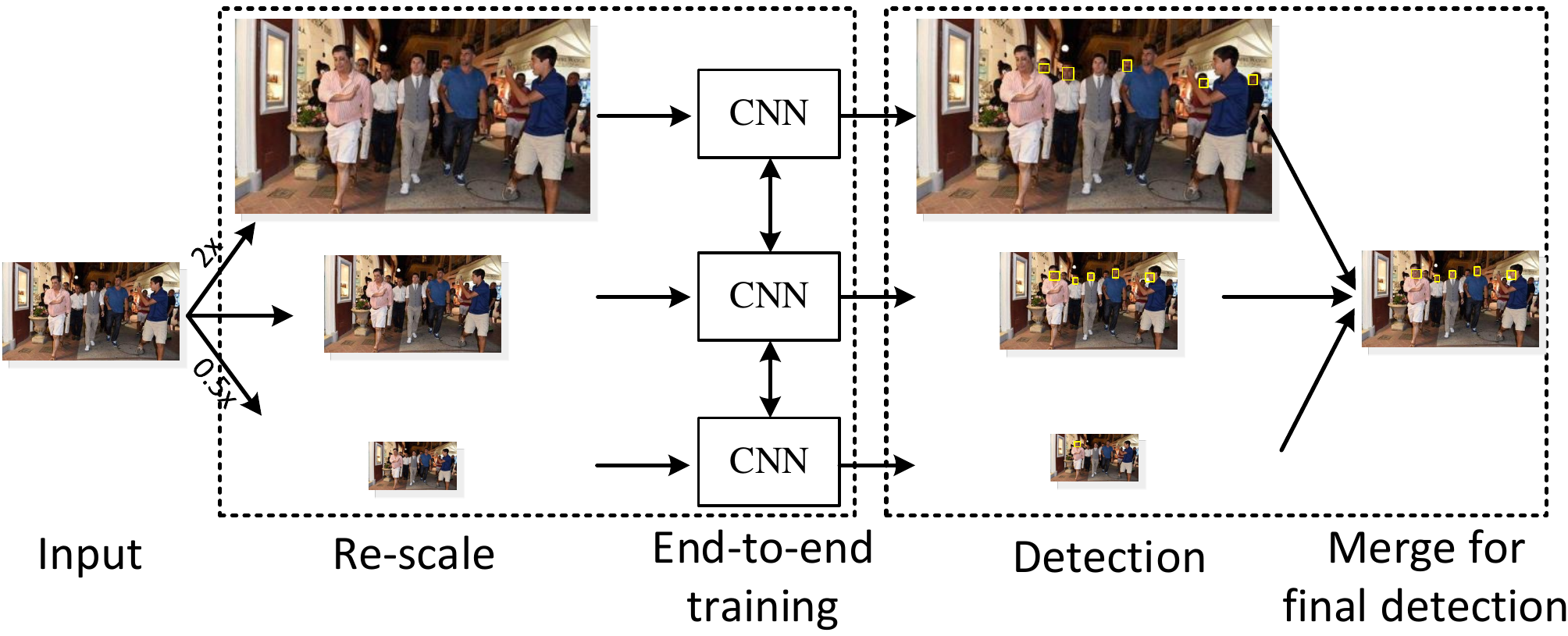}
	\vspace{-0.2cm}
	\caption{The working flowchart of the tiny face detector.}
	\label{fig:tiny_face}
\end{figure}

\textbf{Tiny face detector.}  \hl{The working flowchart of this detector is illustrated in Figure~\ref{fig:tiny_face}.
We first resize each input image into two different resolution images to construct an image pyramid for the training. Those images with different scales/resolutions are inputted into a CNN model to update shared variables. Based on this trained model, we are able to predict the bounding boxes on this image pyramid. Incorporating non-maximum suppression (NMS) method~\cite{neubeck2006efficient} selects and merges all detected bounding boxes to conduct the final detection on the original size of the input image. Note that since a well-trained model has been provided by Hu et al.~\cite{hu2017finding}, we adopt it instead of re-training our own model here.}

\textbf{DR-GAN-based face recognition.} \hl{We employ DR-GAN to recognize user identities, but it is challenging to deal with variations on the faces (e.g., illumination conditions, pose and expression changes). Especially, the pose changes can cause a big drop on face recognition performance. To tackle this challenge, our approach is as follows.}

First, we define face angles that range from -90\textdegree to 90\textdegree. If the face angle equals to 0\textdegree, the face is frontal view, which almost contains all facial information that is positive for recognition. If it is -90\textdegree or 90\textdegree, only one side of face is visible and it is difficult for the model to recognize the face based on a profile. Second, we leverage DR-GAN that can train a network to extract disentangled face representation by fine-tuning on the GAN (Generative Adversarial Networks). It can generate a representation for each face with personal identity information and then the representation can be used for face verification and identification.

\begin{figure}[t]
	\includegraphics[width=0.95\columnwidth]{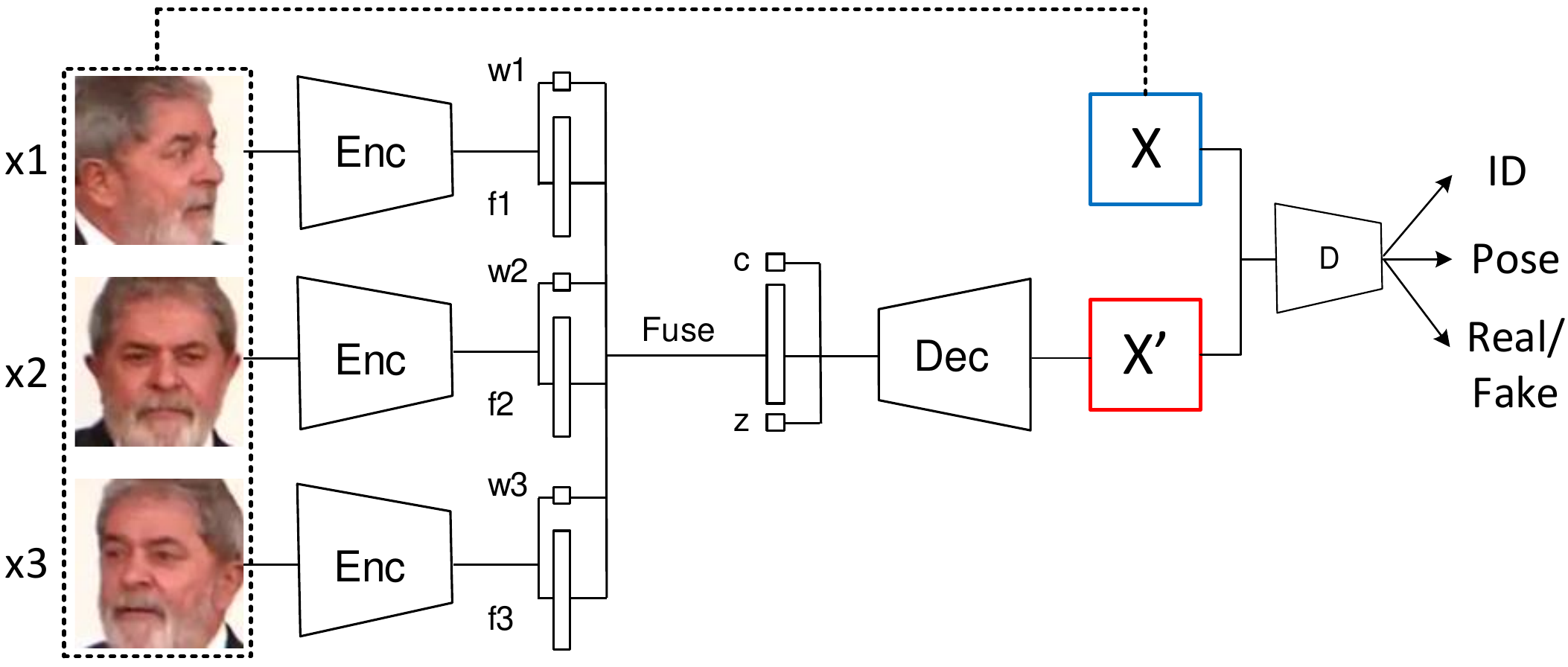}
	\vspace{-0.2cm}
	\caption{Overview of DR-GAN architecture.}
	\label{fig:dr_gan_architecture}
\end{figure}

The DR-GAN face recognition flowchart is shown in Figure~\ref{fig:dr_gan_architecture}. To train a DR-GAN model, several face images of the same identity with different poses are used as the input. Each image will be fed into the encoder, which uses VGG16 as the network structure. Except for generating a 320-dim feature $f$, the encoder also outputs a 1-dim coefficient $w$ for each face. We then compute the fused feature $f'$ by the following equation: 

\begin{equation}\label{eq:dr-gan}
f' =\frac{\sum_{i}^{n}w_i f_i}{\sum_{i}^{n}w_i},
\end{equation}

where $f'$ is a weighed average over all the $f_i$. We next feed $f'$ into a decoder to generate an output image with same size as the input, which is called synthetic image. Note that when feeding $f'$ into the decoder, a pose code $c$ and a random noise $z$ are appended. $c$ can help decoder to generate a synthetic image with arbitrary pose and $z$ can prevent decoder from the issue of over-fitting. Finally, we combine the original input faces with the synthetic image and take them to train a discriminator. When the adversarial training between encoder/decoder and discriminator has converged, we can get an updated encoder. This well-trained encoder is capable of generating the disentangled feature representations for all the input images and we can take those features for face recognition.

\subsubsection{Implementation} We next introduce the implementation of dialing/talking motion detection and user face detection/recognition modules.

\noindent \textbf{Dialing/talking Motion Detection.} We train a binary SVM model for the dialing/talking motion detection. For each training video, \hl{assume 
that the person bounding box} for each frame is known and an HOG descriptor can be extracted for the corresponding bounding box. This applies to both cases of positive and negative videos. The extracted positive and negative HOG descriptors would be the training data for the SVM. In current implementation, we use VLFeat~\cite{VLFeat} to extract HOG descriptors and train the binary SVM.


\noindent \textbf{User Face Detection and Recognition.} We adopt MatConvNet and Tensorflow as the deep learning libraries for the tiny face detector and DR-GAN modules.
First, we develop the tiny face detector and DR-GAN which detect and align the faces for each frame in the testing videos on top of our campus computing servers (MSU HPCC). After that, we feed all the aligned faces into the encoder of the DR-GAN to get the 320-dim features. Besides, we select an aligned frontal face for each person in the testing videos as the gallery image. When conducting the face recognition results, the cosine distance between the feature vector of a detected face in the testing videos and the one from gallery images is computed. If a gallery face has the closest distance with this detected face, the label of this gallery face will be assigned to the detected face.

\subsubsection{Evaluation}
We next evaluate the UCIA system. 

\noindent\textbf{Training data collection.} We recruit 20 volunteers and collect their face images with distinct angles \footnote{Note that all the volunteers' faces are visible.} as well as their dialing/talking actions. These videos are positive training data samples. We use UCF101 database~\cite{soomro2012ucf101} as the negative training samples because they do not contain dialing/talking actions. Moreover, for the detection support of multiple Wi-Fi calling users, we collect videos containing at most 4 different people with different actions over time.

\begin{figure}[t]
	\includegraphics[width=0.95\columnwidth]{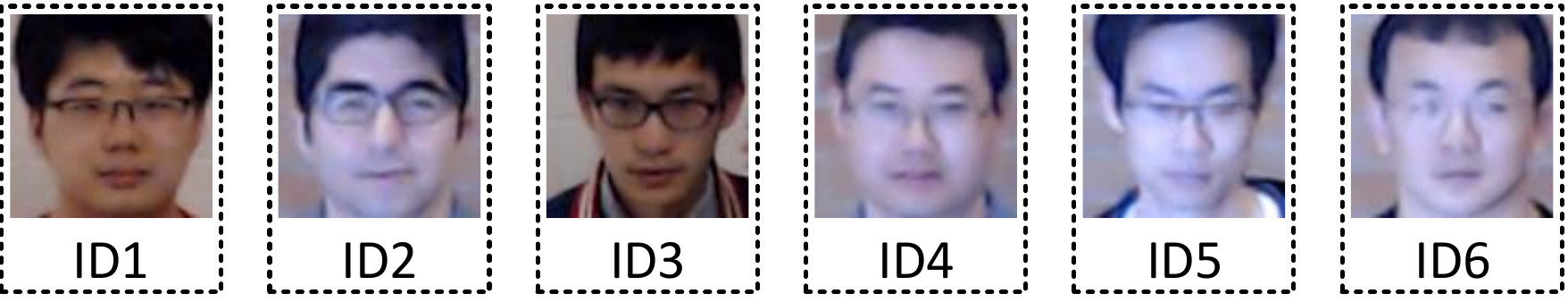}
	\vspace{-0.2cm}
	\caption{Six users participated in the UCIA performance evaluation experiment.}
	\label{fig:person_id}
\end{figure}

\noindent\textbf{Performance evaluation.}
\hl{We evaluate the UCIA system using three performance metrics: (1) motion detection accuracy (MCA), which is computed by Equation~\ref{eq:motion_detection_correctness}; (2) call statistic errors, i.e., time difference between the actual time that a call starts/ends and the estimated time that a call stops/ends; and (3) identity recognition accuracy (IRA), which is computed by Equation~\ref{eq:identity_frame_correctness}.} 
\begin{equation}\label{eq:motion_detection_correctness}
MCA = \hl{\frac{NumCML}{TtlNumMF}},
\end{equation}
\hl{where $NumCML$ is the number of correct motion labels, which are those predicted correctly, and $TtlNumMF$ is the total number of motion frames.}


\begin{equation}\label{eq:identity_frame_correctness}
IRA = \frac{NumCL}{TtlNumF},
\end{equation}
\hl{where $NumCL$ is the total number of frames with correct labels and $TtlNumF$ represents the number of total frames.}

In our experiments, three test scenarios are considered: (1) one person, (2) two persons, and (3) four persons are in the test videos. We use the cross-validation method to evaluate the UCIA performance. We randomly select six participants from our 20 volunteers who contribute training data. They are denoted ID1, ID2, ID3, ID4, ID5, and ID6 as shown in Figure~\ref{fig:person_id}.


\begin{itemize}
  \item \textbf{Single user.} In this experiment, there is only one person recorded in the video. We only ask the participant to dial at least one call. During the experiment, he/she is allowed to make any random actions (e.g., looking at the ground). Note that to simulate the real use scenario, we do not restrict the time period of each Wi-Fi calling call (i.e., it varies with different participants). For each participant, we take five videos; each video lasts 15 to 25 seconds. We use the UICA system to analyze these new videos and the videos recorded from 20 participants.
  When a user picks up his phone and puts it close to his ear in the video, the UCIA is able to detect this action and mark this person using a red bounding box to denote that this person is making a phone call.	Meanwhile, the tiny face detector can detect the face and use the small yellow bounding boxes to mark the face detection results. The DR-GAN assigns an ID label to each detected face (the number shown on the upper-left corner of the bounding box).
  The results show that the UCIA can achieve average accuracy 92.5\% and 96.5\% for motion detection (MCA) and face recognition (IRA), respectively. 
  The call statistic errors are less than 1.5 seconds (i.e., time difference between the estimated time from the UCIA module and the time observed from the actual phone call.). Note that the MCA, the IRA, and the call statistic errors may vary with the sporadic uncontrolled motions of the participants (e.g., they may not always look at the surveillance camera).


\begin{figure}[t]
	\includegraphics[width=0.95\columnwidth]{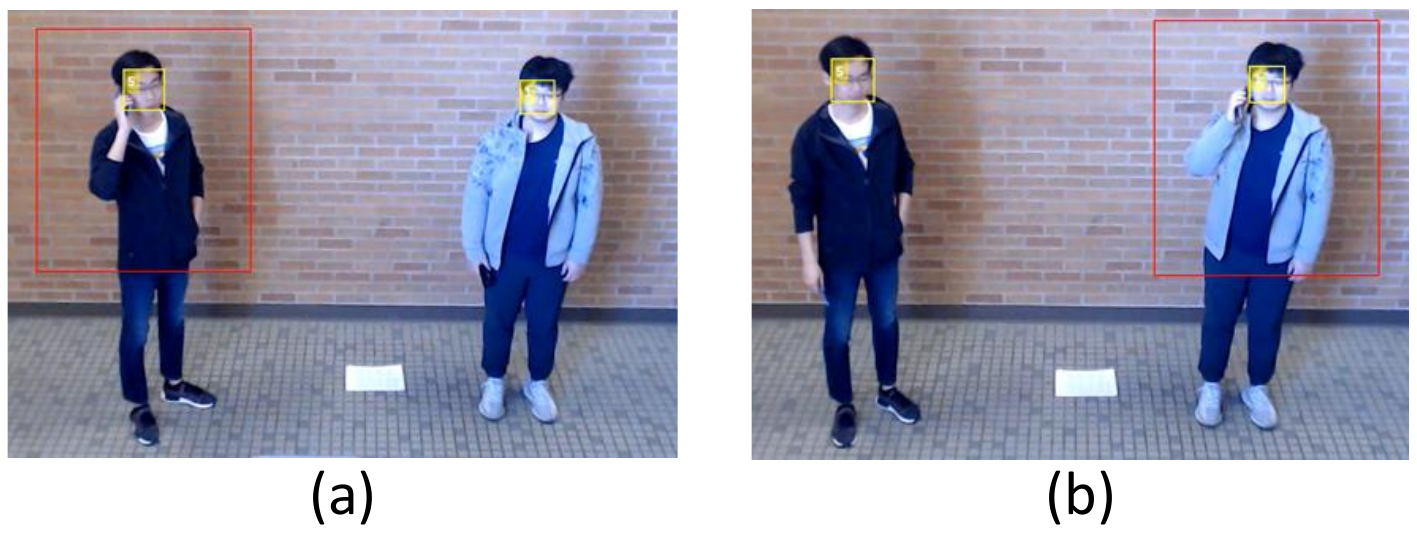}
	\vspace{-0.2cm}
	\caption{Detection and recognition sample results for two people (ID1 (right) and ID5 (left)). The red bounding boxes mark the persons who are making the call. The yellow bounding boxes mark the detection/recognition of the face/identity.}
	\label{fig:detection_recognition_result}
\end{figure}

\begin{table}
	\begin{minipage}{1\columnwidth}
		\begin{center}
			\resizebox{0.85\columnwidth}{!}{
				\scriptsize
				\begin{tabular}{c|c|c|c|c}
					\hline
					\multirow{2}{*}{\textbf{Performance Matrix}} & \multicolumn{2}{c|}{\textbf{ID1}} & \multicolumn{2}{c}{\textbf{ID5}} \\
					\cline{2-5}
					& Mean & Std & Mean & Std \\
					\hline
					\hline
					MDA (Motion Detection Accuracy) & 93.33\%  & 2.38\% & 86.66\% & 6.67\%  \\
					\hline
					IRA (Identity Recognition Accuracy) & 85.87\%  & 3.33\% & 85.21\% & 2.66\% \\
					\hline
					Call conversation time error (sec) & 1.78  & 0.045  & 1.65 & 0.18 \\
					\hline
					Call start time error (sec) & 0.708 & 0.020 & 1.26 & 0.077 \\
					\hline
					Call end time error (sec) & 1.21 & 0.029 & 0.58 & 0.102 \\
					\hline
					\hline
				\end{tabular}
			}
			\vspace{0.1cm}
			\caption{The performance of the UCIA on the 2-user group with ID1 and ID5.}
			\label{tab:2_pedestrians}
		\end{center}
	\end{minipage}
\end{table}

  \item \textbf{Two users.} This experiment includes two participants in the test video. Each user is asked to dial at least one call. They can also do any actions during the experiment. For each two-user group, we conduct 10 experiment runs and record 10 videos. \hl{Figure~\ref{fig:detection_recognition_result} is the snapshots of one of the experiments; there are two users with ID1 and ID5 who walk around or/and make a phone call.  Figure~\ref{fig:detection_recognition_result}(a) and (b) respectively show the actions that user ID5 and user ID1 are making phone calls.
      The mean and standard deviation of the MDA, the IRA, and the call statistic errors are given in Table~\ref{tab:2_pedestrians}.}
      \hl{We have three observations. First, the user ID1 has higher average MDA but lower standard deviation, compared with the user ID5. 
      It may be caused by the user ID5's face outline, which is more blurred.}
      Second, the average MDA and IRA are higher than 85\% for both users ID1 and ID5. Note that our accuracy considers all video frames. In practice, we do not require all frames to be recognized accurately. For example, we can still identify a Wi-Fi user even only one of the frames in the test video is recognized.
      Third, call start/end time errors for both ID1 and ID5 are less than 1.3 seconds, whereas the call conversation time error is around 1.8 seconds. In all of our experiments, the minimum MDA and IRA are higher than 80\% and the maximum call statistic errors are less than 2 seconds.


\begin{figure}[t]
	\includegraphics[width=0.95\columnwidth]{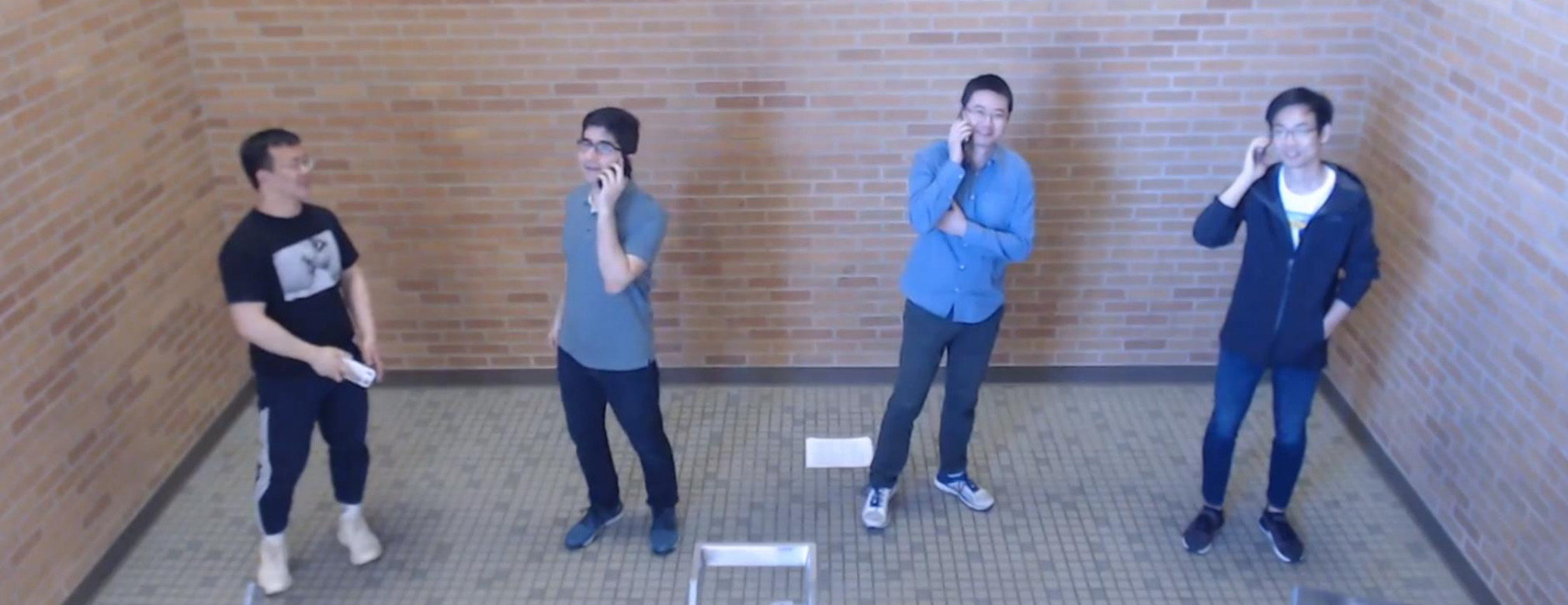}
	\vspace{-0.2cm}
	\caption{One captured frame of the video in a four-user experiment with users ID6, ID2, ID4, and ID5 (from left to right).}
	\label{fig:4_pedestrians}
\end{figure}

\begin{table}
	\begin{minipage}{1\columnwidth}
		\begin{center}
			\resizebox{0.95\columnwidth}{!}{
				\scriptsize
				\begin{tabular}{c|c|c|c|c|c|c|c|c}
					\hline
					\multirow{2}{*}{\textbf{Performance Matrix}} & \multicolumn{2}{c|}{\textbf{ID6}} & \multicolumn{2}{c|}{\textbf{ID2}} & \multicolumn{2}{c|}{\textbf{ID4}} & \multicolumn{2}{c}{\textbf{ID5}} \\
					\cline{2-9}
					& Mean & Std & Mean & Std & Mean & Std & Mean & Std \\
					\hline
					\hline
					MDA (Motion Detection Accuracy) & 95\%  & 0.707\% & 90\% & 2.89\% & 90\% & 1.83\% & 85\% & 1.68\% \\
					\hline
					IRA (Identity Recognition Accuracy) & 95.64\%  & 2.23\%  & 98.09\% & 1.01\% & 92.46\% & 3.33\% & 87.05\% & 4.16\%\\
					\hline
					Call conversation time error (sec) & 1.74 & 0.017 & 2.12 & 0.066 & 1.84 & 0.039 & 2.09 & 0.042\\
					\hline
					Call start time error (sec) & 0.93 & 0.008 & 1.35 & 0.023 & 1.01 & 0.021 & 0.95 & 0.019 \\
					\hline
					Call end time error (sec) & 0.81 & 0.009 & 0.77 & 0.043 & 0.83 & 0.018 & 1.14 & 0.023 \\
					\hline
					\hline
				\end{tabular}
			}
			\vspace{0.1cm}
			\caption{The performance of UCIA on the 4-user group with users ID2, ID4, ID5, and ID6.}
			\label{tab:4_pedestrians}
		\end{center}
	\end{minipage}
\end{table}

  \item \textbf{Four users.} \hl{We here consider four participants in the experiment, which include users ID2, ID4, ID5, and ID6, as shown in Figure~\ref{fig:4_pedestrians}. The experimental settings are the same as the above ones. The mean and standard deviation of the MDA, the IRA, and the call statistic errors are given in Table~\ref{tab:4_pedestrians}.} There are two findings. First, the MDA and the IRA for users ID2, ID4, and ID6 are higher than 90\%, but user ID5 still has relatively low MDA and IRA (i.e., 86.66\% and 85.21\%).
      The call start/end time errors are less than 1.4 seconds, whereas the call conversation time errors are about 2.2 seconds. The results are comparable to those in the 2-user experiment. Second, the performance of UCIA does not downgrade significantly when the number of Wi-Fi calling users doubles. 
\end{itemize}


\noindent\textbf{Summary:} \hl{Our UCIA system can achieve the average accuracy of 80\%~98.5\% for both the MDA and the IRA, and the average call statistic errors below 2.2 seconds, when the number of users varies from one to four. 
We believe that the current performance has been sufficient for the UCIA to identify users and generate useful call statistics, due to two reasons.
First, in practice, it is not required to correctly recognize motions and faces in all video frames. The UCIA is still able to identify a Wi-Fi user even when only one frame of the test video is accurately recognized.
Second, the call statistic errors do not increase with the call lengths, but depend on the accuracy of the motion detection (i.e., moving the phone to ears or putting the phone down). According to a Statista report~\cite{avg-call-length-2013}, the average mobile call length in the U.S. is about 108 seconds. So, the average error takes only a small portion of the average call length, 2.1\% (2.2s/108s).}

\subsection{CS-IP2U (Call Statistic-based IP2User Correlation)}
\label{subsect:cs-ip2u}

\hl{The CS-IP2U correlates user identities with IP addresses based on call statistics extracted by the WiCA and the UCIA.
It mainly considers two kinds of events which are call start and call end.
We denote the happening time of these two events as $TCStart$ and $TCEnd$, respectively.
Ideally, for an identified Wi-Fi calling call, the WiCA outputs $TCStart_{w}$, $TCEnd_{w}$, and $IP$, whereas the UCIA outputs $TCStart_{u}$, $TCEnd_{u}$, and $UserID$. One correlation is identified when $TCStart_{w}==TCStart_{u}$ and $TCEnd_{w}==TCEnd_{u}$. Nevertheless, in practice, it is not the case due to the call statistic errors. 
The CS-IP2U thus considers not only time points but also time intervals. It works as follows.}


\textbf{Step 1.} \hl{We consider a time interval $TCStartInt_{w}$ between $TCStart_{w}-\sigma$ and $TCStart_{w}+\sigma$ and another time interval $TCEndInt_{w}$ between $TCEnd_{w}-\sigma$ and $TCEnd_{w}+\sigma$ for call start and end events, respectively. $\sigma$ is the maximum call statistic error observed in the WiCA (i.e., 1 second).}


\textbf{Step 2.} \hl{A time interval $TCStartInt_{u}$ between $TCStart_{u}-\epsilon$ and $TCStart_{w}+\sigma$ and another time interval $TCEndInt_{w}$ between $TCEnd_{w}-\epsilon$ and $TCEnd_{w}+\epsilon$ are considered for call start and end events, respectively.
$\epsilon$ is the maximum call statistic error observed in the UCIA (i.e., 1.5 seconds).}

\textbf{Step 3.} \hl{We verify two conditions: $TCStartInt_{w} \bigcap TCStartInt_{u} \neq \emptyset$ and $TCEndInt_{w} \bigcap TCEndInt_{u} \neq \emptyset$. If both of them are true, the corresponding $IP$ and $UserID$ are correlated.}

\hl{Note that the current implementation of the CS-IP2U does not support the cases that multiple Wi-Fi calling users start or end calls  near-simultaneously (within the time interval of $Max\{\sigma,\epsilon\}$ (i.e., 1.5 seconds). To address this issue, more fine-grained call statistic information should be extracted by the WiCA and the UCIA. For example, we can infer time periods that users are talking and those that user are listening
by analyzing uplink and downlink Wi-Fi calling voice packets at the WiCA and detecting who are talking~\cite{who-is-talking} at the UCIA. 
We do not implement this advanced feature on the prototype of our proof-of-concept attack, but only demonstrate the feasibility of the correlation between user identities and IP addresses.}

\subsection{Evaluation on the UPIS System} 

\hl{We evaluate the performance of the UPIS system, which integrates the WiCA, the UCIA, and the CS-IP2U. The CS-IP2U requires to associate time and events between the WiCA and the surveillance camera, so the clock synchronization between them is needed. In our implementation, they are deployed at the same laptop; otherwise, the precision time protocol (PTP)~\cite{ptp-protocol} can be used for the synchronization.
We here repeat the 2-user and 4-user experiments as conducted in Section~\ref{sect:module_uca}, and collect and analyze the Wi-Fi calling packets at the WiCA. In all the experiments, the users are allowed to dial more than one calls and do any random actions.
We have 5 runs for each experiment and each participant dials 10-20 calls.}



\hl{We gauge the average errors of the call start/end times estimated by the WiCA/UCIA by comparing them with the actual ones, and the success rate of identifying the association between user identities and the IP addresses. The success rate is the ratio of the number of successful identification times to the total number of Wi-Fi calling calls for each user.} 
\hl{The evaluation results are given in Table~\ref{tab:pim_performance}. We have two observations. First, the average errors of the call start/end times are smaller than 1.52 seconds, which are similar to the evaluation results of individual WiCA and UCIA modules. Second, there is a very small percentage that the CS-IP2U module does not successfully identify the identity/IP association (i.e., 97.4\%, 74/76).
For example, it fails to identify the identity/IP associations for ID5 and ID6 in a 4-user experiment.
The root cause is that these two users dial and end their calls within a small 1.5-second interval.
This issue calls for more fine-grained call statistic information which we discussed in Section~\ref{subsect:cs-ip2u}.}

\begin{table}
	\begin{minipage}{1\columnwidth}
		\begin{center}
			\resizebox{0.95\columnwidth}{!}{
				\scriptsize
				\begin{tabular}{c|c|c|c|c|c|c}
					\hline
					\multirow{2}{*}{\textbf{Performance Matrix}} & \multicolumn{2}{c|}{\textbf{Two Users}} & \multicolumn{4}{c}{\textbf{Four Users}}  \\
					\cline{2-7}
					& ID1 & ID5 & ID6 & ID2 & ID4 & ID5 \\
					\hline
					\hline
					WiCA start time error (sec) & 0.11 & 0.08 & 0.25 & 0.55 & 0.15 & 0.08 \\
					\hline
					WiCA end time error (sec) & 0.09 & 0.18 & 0.17 & 0.23 & 0.37 & 0.10 \\
					\hline
					UCIA start time error (sec) & 0.91 & 1.21 & 1.51 & 1.34 & 0.53 & 0.98  \\
					\hline
					UCIA end time error (sec) & 0.69 & 0.83 & 0.62 & 0.99 & 0.76 & 1.19  \\
					\hline
                    \hline
					\textbf{Success rate of identifying User ID and IP} & 100\% (15/15) & 100\% (17/17) & 95\% (19/20) & 100\% (19/19) & 100\% (17/17) & 95\% (19/20)\\
					\hline
					\hline
				\end{tabular}
			}
			\vspace{0.1cm}
			\caption{Overall performance of the UPIS system.} 
			\label{tab:pim_performance}
		\end{center}
	\end{minipage}
\end{table}

\subsection{Real-world Impact}
To the best of our knowledge, the UPIS is the first system which can identify IP addresses, call statistics, and the identities of the Wi-Fi calling users. We believe that \hl{some use scenarios can benefit from the UPIS} in practice. For example, in order to be against possible campus assaults, universities can use it to infer students' personality (e.g., conscientiousness~\cite{de2013predicting}), recent mood (e.g., stressful~\cite{thomee2011mobile}), malicious behaviors (e.g., dialing spamming calls)~\cite{DBLP:conf/ceas/BalasubramaniyanAP07}, or the network services and websites they surf.
The UPIS can be also deployed at airports to be against terrorists. It allows the law enforcement agents to identify suspects' phone models and IP addresses, and further remotely install the malware on their phones for monitoring. The remote installation can be achieved by exploiting public security vulnerabilities of the target devices. 
Note that we do not advocate any use scenarios compromising user privacy no matter whether their purposes are benign or not.

	\section{Telephony Harassment/Denial of Voice Service (THDoS) Attack}
\label{sect:thdos}

\hl{We next devise an attack that causes telephony harassment or denial of voice service on Wi-Fi calling users.
It can bypass the security defenses deployed on Wi-Fi calling devices and the infrastructure. 
It manipulates the delivery of the Wi-Fi calling signaling and voice packets, and has several variances,
e.g., hiding the callee's alerting, extra incoming calls, mute calls, etc.
In the following, we present attack designs and real-world negative impacts after introducing the Wi-Fi calling call flow.}

\begin{figure}[t]
	\centering
	\includegraphics[width=0.8\columnwidth]{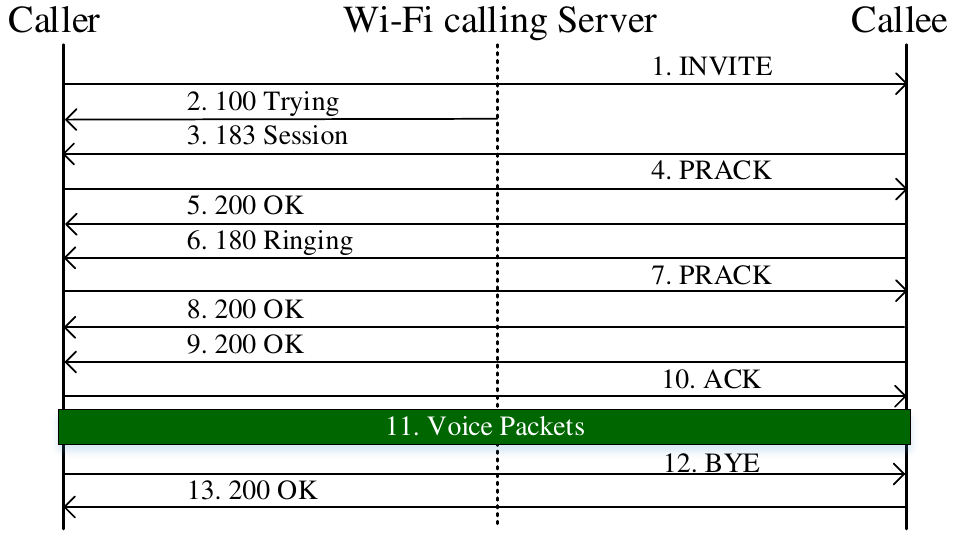}
	\vspace{-0.4cm}
	\caption{Wi-Fi calling call flow diagram.}
	\label{fig:sip_flow}
\end{figure}

\subsection{Wi-Fi calling Call Flow}
The signaling of the Wi-Fi calling call relies on the session initiation protocol (SIP).
Figure~\ref{fig:sip_flow} shows an outgoing call flow of the Wi-Fi calling.
To initiate a call, the caller sends an SIP \texttt{INVITE} message, which specifies the capabilities (e.g., voice codec) of the caller, to the callee.
Then, the Wi-Fi calling server at the IMS system replies an \texttt{100 Trying} message to indicate that the call setup is in progress. In the meantime, the callee replies a list of chosen voice codecs in an \texttt{183 Session} message to the caller.
Afterwards, the caller sends a \texttt{PRACK} (Provisional Acknowledge) message to inform the callee about the selected codec.
After the callee's phone rings, it will send a \texttt{180 Ringing} message to the caller and then the caller's phone rings.
Once the callee answers the call, two ends start to chat after exchanging \texttt{200 OK} and \texttt{ACK} messages. 
A \texttt{BYE} message is finally sent from the end who terminates the call and the other end acknowledges it with a \texttt{200 OK} message.

\subsection{Attack Design}
We launch an ARP spoofing attack against a Wi-Fi calling user and then intercept all of the user's Wi-Fi calling packets. Based on the identified traffic characteristics of Wi-Fi calling calls in Section~\ref{sect:module_wica}, we are able to identify specific signaling and voice packets.
We discover that there are four attack variances by discarding various Wi-Fi calling packets with different patterns.
Table~\ref{tab:attack1_result} summarizes the results where we drop different signaling packets or voice packets of an outgoing call.

\begin{table}[t]
	\resizebox{0.9\columnwidth}{!}{
		\scriptsize
		\begin{tabular}{|>{\centering\arraybackslash}p{0.2cm}|p{2cm}|p{1.2cm}|p{5cm}|}
			\hline
			\multirow{2}{*}{\textbf{No.}} & \multirow{2}{*}{\parbox{2cm}{\textbf{Dropped Packets}}} & \multirow{2}{*}{\textbf{Sender}} & \multirow{2}{*}{\textbf{Results}}\\
            &&&\\
			\hline
			\hline
			\multirow{1.2}{*}{1} & \multirow{1.2}{*}{INVITE} & \multirow{1}{*}{Caller} & \multirow{1.2}{*}{\parbox{4cm}{Caller initiates a cellular-based call.}} \\
			\hline
			\multirow{1.2}{*}{2} & \multirow{1.2}{*}{100 Trying} & \multirow{1.2}{*}{Server} & \multirow{1.2}{*}{No effect.} \\
			\hline
			\multirow{1.2}{*}{3} & \multirow{1.2}{*}{183 Session} & \multirow{1.2}{*}{Callee} & \multirow{1.2}{*}{\parbox{4cm}{Two outgoing calls arrive at callee.}} \\
			\hline
			\multirow{1.2}{*}{4} & \multirow{1.2}{*}{PRACK} & \multirow{1.2}{*}{Caller} & \multirow{1.2}{*}{No effect.} \\
			\hline
			\multirow{1.2}{*}{5} & \multirow{1.2}{*}{200 OK} & \multirow{1.2}{*}{Callee} & \multirow{1.2}{*}{No effect.} \\
			\hline
			\multirow{2}{*}{6} & \multirow{2}{*}{180 Ringing} & \multirow{2}{*}{Callee} & \multirow{2}{*}{\parbox{5cm}{Caller will not enter conservation state. His/her phone gets stuck in the dialing screen.}} \\
			&&& \\
			
			\hline
			\multirow{1.2}{*}{7} & \multirow{1.2}{*}{PRACK} & \multirow{1.2}{*}{Caller} & \multirow{1.2}{*}{No effect.} \\
			\hline
			\multirow{1.2}{*}{8} & \multirow{1.2}{*}{200 OK} & \multirow{1.2}{*}{Callee} & \multirow{1.2}{*}{\parbox{4cm}{Caller keeps hearing the alerting tone.}} \\
			\hline
			\multirow{1.2}{*}{9} & \multirow{1.2}{*}{200 OK} & \multirow{1.2}{*}{Callee} & \multirow{1.2}{*}{\parbox{4cm}{Caller keeps hearing the alerting tone.}}  \\
			\hline
			\multirow{1.2}{*}{10} & \multirow{1.2}{*}{ACK} & \multirow{1.2}{*}{Caller} & \multirow{1.2}{*}{No effect.} \\
			\hline
			\multirow{1}{*}{11} & \multirow{1}{*}{\parbox{2cm}{Voice Packets}} & \multirow{1}{*}{\parbox{0.8cm}{Caller/Callee}} & \multirow{1}{*}{\parbox{4cm}{Call drops or voice quality downgrades.}}\\
			
			\hline
			\multirow{2}{*}{12} & \multirow{2}{*}{BYE} & \multirow{2}{*}{Caller} & \multirow{2}{*}{\parbox{5cm}{Callee gets stuck in the conversation state for 20s. Afterwards, the call is terminated.}}\\
			&&& \\
			
			\hline
			\multirow{1.2}{*}{13} & \multirow{1.2}{*}{200 OK} & \multirow{1.2}{*}{Callee} & \multirow{1.2}{*}{No effect.}\\
			\hline			
		\end{tabular}
	}
	\vspace{0.1cm}
	\caption{The results where we drop different Wi-Fi calling signaling packets or voice packets of an outgoing call.}
	\label{tab:attack1_result}
\end{table}

\smallskip
\textbf{Annoying-Incoming-Call Attack.} \hl{The callee as the victim would receive multiple incoming calls from the caller.}  There are two approaches. First, the adversary drops the \texttt{183 Session Progress} message sent by the callee, and then the caller's Wi-Fi calling device would initiate another VoLTE call towards the callee. Second, the adversary discards the \texttt{180 Ringing} message sent by the callee, and then it would cause the caller's Wi-Fi calling device to get stuck in the dialing screen. The caller does not hear any alerting tone, but the callee's device would ring. The caller may thus keep redialling.

\smallskip
\textbf{Zombie-Call Attack.} \hl{The caller's device can be forced to get stuck in the dialing screen, when the adversary discards the \texttt{200 OK} message sent by the callee. The message indicates that the call has been answered. So the caller's device gets stuck in the dialing screen and keeps hearing the alerting tone. The call conversation is thus never started.} 

\smallskip
\textbf{Mute Call Attack.} Two parties of a Wi-Fi calling call are both victims. This attack does not aim to terminate the call but only mute the victims' speech. Our result shows that the adversary can mute the call up to 8 seconds by dropping voice packets. Note that if the voice suspension time is longer than 8 seconds, the voice call will be terminated by the network.

\begin{table}[t]
    \center
	\resizebox{0.6\columnwidth}{!}{
		\scriptsize
		\begin{tabular}{c|c}
			\hline
			\textbf{Drop Rate (\%)} & \textbf{Voice Quality} \\
			\hline
			\hline
			below 20\% & No clear impact.  \\
			\hline
			40-60\% & Some noises.  \\
			\hline
			70-90\% & Conversation is hardly continued. \\
			\hline
			100\% & Call is terminated by the network. \\
			\hline
			
		\end{tabular}
	}
	\caption{Voice quality varies with the drop rate of voice packets.}
	\label{tab:attack1_voice}
\end{table}

\smallskip
\textbf{Telephony Denial-of-Voice-Service Attack.} Both the caller and the callee are victims. This attack aims to downgrade the voice quality of a Wi-Fi calling call so that the conversation is hard to be continued, whereas the deployed system-switch security mechanism (i.e., switching a Wi-Fi calling call to cellular network voice call) is suppressed.
It is achieved by controlling the drop rate of the intercepted Wi-Fi calling packets of the victim. Table~\ref{tab:attack1_voice} shows the negative impact on voice quality for different drop rates. There are four findings. First, when the packet drop rate is below 20\%, users do not report voice quality downgrade.  Second, when the packet drop rate is increased to 40\%-60\%, some of users may notice some noises. Third, when the packet drop rate is 70\%-90\%, the voice call is hardly continued. Fourth, when the packet drop rate is 100\%, the Wi-Fi calling call will be terminated in less than 10 seconds.

\subsection{Real-world Impact}
\label{subsect:feas-wifi-calling-attacks}
The real-world impact of our THDoS attack can be significant in practice. Most of U.S. universities have deployed campus Wi-Fi networks. 
However, our studies show that the campus Wi-Fi is the best attack surface for adversaries. Take Michigan State University as an example. Its campus Wi-Fi (MSUNet) provides all students, the faculty, and the staff with free Wi-Fi access. In our 2-min experiment, we discover that more than 700 devices including smartphones, tablets, and computers, connect to MSUNet. All the devices are served by the same gateway which is vulnerable to ARP spoofing attack. Note that we validated the gateway's vulnerability through our own devices. As a result, adversaries are likely to intercept the packets of all Wi-Fi calling devices and launch the THDoS attack. Note that the campus Wi-Fi at Michigan State University is not the only Wi-Fi infrastructure with this issue. We find that it also exists in many universities' campus Wi-Fi, such as New York University, University of California Berkeley, Northeastern University, etc.

	\section{Recommended Solutions}
\label{sect:sol}

\hl{In this section, we propose not only a short-term solution using virtual private network (VPN), which can mitigate the threats quickly in practice, but also a long-term solution upgrading Wi-Fi Calling standards, which can address the security vulnerabilities thoroughly. 
Though the ARP spoofing detection is a straightforward solution for the aforementioned threats, the cost is too high to be deployed in large-scale networks.
Specifically, for a public Wi-Fi network, it requires to maintain a database for all the users.
We thus skip this solution option in this section.}


\subsection{Using VPN Service}
\hl{Currently, most of the IPSec packets sent between the Wi-Fi calling devices and the ePDG belong to the Wi-Fi calling services. It allows the adversary to easily identify the Wi-Fi calling events from encrypted IPSec packets. 
We suggest that operators can add noise to the transmitted data and then leverage VPN services (e.g., AT\&T IPSec VPN~\cite{att-business-vpn} and IPVanish VPN) to carry all the data, which can be from Wi-Fi calling services, other applications, or/and noise.
It can increase the difficulties of the inference on the Wi-Fi calling packets, and it is much more challenging to
develop a general cross-device/cross-user/cross-operator approach for the inference. 
Note that Wi-Fi calling users may not need to pay extra cost since some popular VPN services (e.g., NordVPN, TunnelBear VPN and Golden Frog VyprVPN)~\cite{best-vpn-2018} provide free versions and many commercial off-the-shelf home Wi-Fi router manufacturers (e.g., Netgear and TP-Link) support VPN server functions on their Wi-Fi routers.}

\hl{To confirm the viability of this solution,
we set up a VPN channel on our Wi-Fi calling devices and run various applications simultaneously while using the Wi-Fi calling service.
We run three popular applications, Google Gmail, Tencent Wechat, and Google Maps, in the background.
For different combinations of them,
we conduct 10 experiments and directly apply the previous trained model to them.
Table~\ref{tab:sol_tab1} shows the performance of this short-term VPN solution. 
It is observed that when the traffic of the running applications mixes with that of the Wi-Fi calling, the detection correctness largely downgrades.
Though this quick remedy does not completely eliminate security threats, we believe that it can significantly reduce the real-world damages caused by the Wi-Fi calling based attacks in a short time.
}


\begin{table}[t]
	\begin{minipage}{0.7\columnwidth}
	\center
	\resizebox{1\columnwidth}{!}{
		\scriptsize
		\begin{tabular}{c|c}
			\hline
			\textbf{Running Application(s) \footnote{Excluding the Wi-Fi calling services.}} & \textbf{Correctness of Detection} \\
			\hline
			\hline
			0 & 100\%  \\
			\hline
			1 & 10\%  \\
			\hline
			2 & 0\% \\
			\hline
			3 & 0\% \\
			\hline	
		\end{tabular}
	}
	\vspace{0.1cm}
	\caption{Detection correctness when applying the previous trained model to the traffic on the VPN tunnel.}
	\label{tab:sol_tab1}
\end{minipage}
\end{table}


\subsection{Upgrading Wi-Fi Calling Standards}

\hl{The traffic of the Wi-Fi calling service may traverse public, insecure Wi-Fi networks, so its security shall be at least one of the top-priority features in the standards. We here propose two mechanisms for the need of the standard upgrade. 
Note that we aim to take security actions against malicious Wi-Fi networks and possible corresponding attacks, but not secure public/private Wi-Fi networks that are out of the operator's control.}


\hl{First, both of the Wi-Fi calling devices and the infrastructure should detect whether Wi-Fi calling users are suffering from attacks.
It can be done based on the Wi-Fi calling service quality.
Once detected, the Wi-Fi calling users should switch their calls to another Wi-Fi networks or the conventional cellular network.
It requires SRVCC/DRVCC to be upgraded to add the service quality to be one new trigger factor of network switches.}


\hl{Second, both of Wi-Fi calling infrastructure and devices should deploy some security defenses against the well-known Wi-Fi based attacks. For example, the infrastructure should update the RAN and ANDSF rules to exclude malicious networks from the root. The Wi-Fi calling devices should deploy security defenses (e.g., ARP Guard -WiFi security~\cite{arpguard}) to detect or defend against several common attacks (e.g., the ARP spoofing attack) to ensure that the Wi-Fi networks to which they connect are secure, since not all the attacks can be detected by the infrastructure. } 


\hl{Note that this long-term solution can radically address V1, V2, and V3. Moreover, when those three vulnerabilities are eliminated, V4 does not exist.}

	\section{Related Work}
\label{sect:related}

\smallskip
\noindent
\textbf{Cellular Network Security.} Cellular network security is getting more attention in recent years. Christian et al.~\cite{peeters18} proposed
Sonar to detect SS7 redirection attacks with audio-based distance bounding.
Reaves et al.~\cite{reaves2017authenticall} introduced AuthentiCall to protect voice calls made over traditional telephone networks by leveraging now-common data connections available to call endpoints.
Another study~\cite{reaves2016sending} analyzed nearly 400,000 text messages sent to public online SMS gateways over the course of 14 months and offered insights into the prevalence of SMS spam and behaviors. 
\hl{The other three works\cite{DBLP:journals/comsur/GeneiatakisDKLGES06, Geneiatakis05sipsecurity, rehman2014security} study various attacks for SIP on different levels, discuss a potential attack based on SIP signaling, and classify existing SIP attacks and defenses, respectively.}
Compared with them, our work focuses on the security of the newly deployed Wi-Fi calling service security, which has not been fully explored yet.

\smallskip
\noindent
\textbf{VoIP and VoLTE Security.}
The security problem of the VoIP and VoLTE system has attracted lots of attentions. Two studies~\cite{compagno2017don,fang2016voice} examine side-channel attacks on VoIP traffic. McGann et al.~\cite{mcgann2005analysis} analyzed the security threats and tools in the VoIP system. Several security issues (e.g., Toll Fraud) of VoIP applications were discussed in~\cite{hung2006security}. Li et al.~\cite{li2015insecurity} examined the security implications of VoLTE, which include several vulnerabilities (e.g., improper charing policies).
Dacosta et al.~\cite{dacosta2009improving} proposed the use of a modified version of OpenSER to improve authentication performance of distributed SIP proxies.
This paper studies the Wi-Fi calling service \hl{from the perspectives of the standard, the implementation, and the operation,} which are not covered by the prior arts.

\smallskip
\noindent
\textbf{Side-Channel Attacks Against Mobile Systems.}  The side-channel information leakage against mobile systems has been a popular research area in recent years. Current studies~\cite{compagno2017don,fang2016voice} target the side-channel information leaked by mobile users' traffic, which is generated by some particular Internet services, and then seek to infer users' activities. 
The work~\cite{bocchi2016magma} introduces the analysis on automatic fingerprinting of mobile applications for arbitrarily small samples of Internet traffic. Ali et al.~\cite{ali2018toward} illustrated that each app leaves a fingerprint on its traffic behavior (e.g., transfer rates, packet exchanges, and data movement).
Another work~\cite{taylor2018robust} demonstrates automatic fingerprinting and real-time identification of Android applications from their encrypted network traffic, which even could work when HTTPS/TLS is employed. Eskandari et al.~\cite{eskandari2017analyzing} analyzed the personal data transfers in mobile apps and revealed that 51\% of these apps did not provide any privacy
policy. The paper~\cite{block2018my} demonstrates discerning of mobile user location within commercial GPS resolution by leveraging the ability of mobile device magnetometers to detect externally generated signals in a permissionless attack. Reaves et al.~\cite{reaves2017mo} did the security analysis on the branchless banking applications.
Different from them, we focus on the insecurity of the cellular Wi-Fi calling service, which is stipulated by the 3GPP and is going to be deployed globally on billions of mobile devices in the near future.

\smallskip
\noindent
\textbf{Wi-Fi Security.} There are many novel studies related to Wi-Fi security. Liu et al.~\cite{liu2018authenticating} used the fine-grained channel information to authenticate the user. Lee et al.~\cite{lee2014run} examined the limitations of the existing jamming schemes against channel hopping Wi-Fi devices in dense networks. Li et al.~\cite{li2016demographics} inferred user demographic information by exploiting the meta-data of Wi-Fi traffic. Another study~\cite{consolvo2010wi} proposes the system, the Wi-Fi Privacy Ticker, to improve participants' awareness of the circumstances in which their personal information is transmitted. \hl{However, we here focus on a cellular network technology, Wi-Fi calling, the security of which is different from that of conventional Wi-Fi networks.} 

\smallskip
\noindent
\textbf{Wi-Fi Calling Security.} Wi-Fi calling security is a new research area and has not been fully studied by the academic yet, since carriers just deployed their Wi-Fi calling services in recent years. 
Current researchers mainly focus on the security vulnerabilities on Wi-Fi calling devices. Specifically, Beekman et. al pointed out that T-Mobile Wi-Fi calling devices (e.g., Samsung S2) are vulnerable to invalid server certificates~\cite{beekman2013man}. Chalakkal et. al studied SIM-related security issues on Wi-Fi calling devices~\cite{chalakkal2017practical}. However, our work examines the Wi-Fi calling security from all the three aspects: standards, operations and implementations.

	\section{conclusion}
\label{sect:concl}

The Wi-Fi calling service is thriving and being deployed worldwide. In this work, we conduct the first study on the security implication of the operational Wi-Fi calling service over three major U.S. operators and state-of-the-art Wi-Fi calling devices (e.g., Google Nexus 6P, Apple iPhone 8, Samsung Galaxy S8). We discover four security vulnerabilities which stem from the design defects of Wi-Fi calling standards (V1 and V4), operational slips (V3) of operators, and implementation issues (V2) of Wi-Fi calling devices. 

\hl{By} exploiting the vulnerabilities and the state-of-the-art computer visual recognition technologies, adversaries are able to infer the Wi-Fi calling user's privacy and launch the telephony harassment or denial of voice service attack. In the attack of user privacy leakage, adversaries can infer user identity, call statistics, device information, personalities, mood, malicious behaviors, etc. In the attack of telephony harassment/DoS attack, the adversaries are able to shut the essential voice/text services down on the victims' smartphones while the security defenses deployed by Wi-Fi calling service providers and device manufacturers are suppressed. 

The fundamental issue is that the \hl{conventional} security defenses well examined in cellular network services are simply applied to the Wi-Fi calling service without considering its specific security threats. We thus propose two remedies to alleviate real-world damages after \hl{getting to the root of the vulnerabilities.} 
The ultimate solution calls for a concerted effort among all parties involved.

The Wi-Fi calling service is still at its early rollout, so the lessons learned from three major U.S. carriers can help secure mobile ecosystem and facilitate the global deployment,
as well as provide new design insights for upcoming 5G networks. We hope that our initial study will stimulate more research efforts on the Wi-Fi calling service from both
academia and industry.

	\section{Updates from Industry}
\label{sect:update}

To secure a number of Wi-Fi calling service users, we are contacting tested Wi-Fi calling service operators (e.g., Verizon, AT\&T, and T-Mobile) and device manufacturers (e.g., Google, Samsung, and Apple), reporting our findings to them, and waiting for their responses. Currently, the Google Android security team has conducted an initial severity assessment on our report (i.e., DoS attack against essential voice service on Nexus 6P ). They confirmed the security issues we reported, classified them into the Low severity category, and promised that the issues would be fixed at the next appropriate opportunity. Note that Google's published severity assessment matrix has strict rules to determine the severity of the security issues they received. For example, a local permanent denial of service attack (device requires a factory reset to recover) or remote temporary device denial of service (cause devices to remotely hang or reboot) can be only considered a moderate severity issue (i.e., just one level higher than Low severity). However, we do appreciate Google's Android security team is willing to verify our findings and release remedies to its users in a short time. We hope that other Wi-Fi calling device manufacturers and service operators can take similar actions as Google did.

	\begin{acks}
		
    We greatly appreciate Xinyu Lei, Wen Zhong, Jiawei Li, et al. who participated in the experiments of Wi-Fi calling user privacy inference system. This work is supported in part by the National Science Foundation under Grants No. CNS-1814551 and CNS-1815636. Any opinions, findings, and conclusions or recommendations expressed in this material are those of the authors only and do not necessarily reflect those of the National Science Foundation.
		
	\end{acks}
	
	\bibliographystyle{ACM-Reference-Format}
	\bibliography{bib/17hotmobile,bib/18CNS,bib/18CNS_tpc,bib/visual,bib/18TOPS}

\end{document}